\documentclass[letterpaper,twocolumn,10pt]{article}
\usepackage{usenix-2020-09}

\usepackage{xspace}
\newcommand{\sysname}{{\tt kTrans}\xspace}

\usepackage{enumitem}
\usepackage{booktabs}

\usepackage{pifont} 
\newcommand{\cmark}{\ding{51}}%
\newcommand{\xmark}{\ding{55}}%

\usepackage{caption}
\usepackage{subcaption}

\usepackage{tikz}
\usepackage{amsmath}
\usepackage{verbatim}

\usepackage{amssymb}

\usepackage{multirow}
\usepackage{multicol}

\begin{document}

\date{}

\title{\Large \bf \sysname: Knowledge-Aware Transformer for Binary Code Embedding}

\author{
{\rm Wenyu Zhu}\\
Tsinghua University
\and
{\rm Hao Wang}\\
Tsinghua University
\and
{\rm Yuchen Zhou}\\
Beijing University of Technology
\and
{\rm Jiaming Wang}\\
Huazhong University of Science and Technology
\and
{\rm Zihan Sha}\\
Information Engineering University
\and
{\rm Zeyu Gao}\\
Tsinghua University
\and
{\rm Chao Zhang}\\
Tsinghua University
} 


\maketitle

\begin{abstract}
Binary Code Embedding (BCE) has important applications in various reverse engineering tasks such as binary code similarity detection, type recovery, control-flow recovery and data-flow analysis. Recent studies have shown that the Transformer model can comprehend the semantics of binary code to support downstream tasks. However, existing models overlooked the prior knowledge of assembly language. In this paper, we propose a novel Transformer-based approach, namely \sysname, to generate knowledge-aware binary code embedding. By feeding \textit{explicit knowledge} as additional inputs to the Transformer, and fusing \textit{implicit knowledge} with a novel pre-training task, \sysname provides a new perspective to incorporating domain knowledge into a Transformer framework. We inspect the generated embeddings with outlier detection and visualization, and also apply \sysname to 3 downstream tasks: Binary Code Similarity Detection (BCSD), Function Type Recovery (FTR) and Indirect Call Recognition (ICR). Evaluation results show that \sysname can generate high-quality binary code embeddings, and outperforms state-of-the-art (SOTA) approaches on downstream tasks by 5.2\%, 6.8\%, and 12.6\% respectively. \sysname is publicly available at: 
\url{https://github.com/Learner0x5a/kTrans-release}

\end{abstract}
\section{Introduction} \label{sec:intro}



Binary code embedding (BCE), also known as binary code representation learning, aims to map unstructured binary code into a low-dimensional space where the binary code is represented as embeddings. Through binary code embedding, numerous traditional binary code analysis tasks can be improved using deep learning methods, such as binary code similarity detection~\cite{liu2018alphadiff,li2021palmtree,wang2022jtrans,artuso2022binbert,xu2017neural,lee2017instruction2vec,ding2019asm2vec,massarelli2019safe,zhang2022combo}, function boundary recognition~\cite{shin2015boundary,pei2020xda}, function signature recovery~\cite{chua2017type}, value set analysis~\cite{guo2019deepvsa}, and indirect call identification~\cite{zhu2023callee}, etc.
With the development of large language models, the Transformer~\cite{devlin2018bert} has been proven to be an efficient language representation model and has also achieved excellent performance in various binary code analysis tasks.


Existing binary code embedding methods can be roughly divided into two categories: manual embedding and learning-based embedding.
Manual embedding involves representing binary code with manually constructed numerical features, such as Gemini~\cite{xu2017neural}, Instruction2Vec~\cite{lee2017instruction2vec}, etc. However, such methods require extensive domain expertise and are often task-specific, resulting in poor transferability.
On the other hand, learning-based embedding automates the process of generating embeddings for binary code using representation learning methods, such as Asm2Vec~\cite{ding2019asm2vec}, SAFE~\cite{massarelli2019safe}, etc. These methods learn features automatically from data, avoiding biases introduced by manual feature engineering.
Furthermore, with the widespread application of Transformers, many Transformer-based approaches have been proposed and achieved state-of-the-art performance on various downstream tasks, such as PalmTree~\cite{li2021palmtree}, jTrans~\cite{wang2022jtrans}, COMBO~\cite{zhang2022combo}, etc.
Despite the success of existing approaches in binary code analysis tasks, they still have several limitations.

Firstly, existing approaches lack the utilization of prior knowledge inherent in binary code.
Most approaches treat binary code as natural language and directly apply natural language models to assembly language~\cite{artuso2022binbert,wang2022jtrans}. However, assembly language contains knowledge about instruction set architectures (ISAs), including the types of instruction opcodes, operand types, relationships between opcodes, and other information.
For example, for \texttt{rax, eax, ax, al}, a natural language model would consider them as independent tokens, but in fact, the latter three are all part of \texttt{rax}.
Therefore, if the model could understand the relationship between \texttt{eax} and \texttt{al}, it would be able to capture the data-flow relationship between instructions when modeling sequences such as \texttt{movzx eax, byte [r8]; cmp byte [rdx], al}.

Secondly, existing approaches lack understanding of instructions and thus are unable to model program execution behavior.
Instructions are the fundamental units of program execution, but most approaches directly apply natural language models to assembly language without a clear understanding of instruction boundaries. For example, PalmTree requires users to provide instruction boundaries, while BinBert~\cite{artuso2022binbert} lacks information about instruction boundaries altogether.
Without a proper understanding of instructions, the model may not differentiate between \texttt{['pop','rbp']} and \texttt{['rbp','pop']}.

Moreover, existing approaches lack modeling of implicit dependencies in binary code. Assembly instructions are not independent entities but have implicit dependencies, such as the global flag register EFLAGS.
Current approaches partially address implicit dependencies through manual design. For example, jTrans models instruction jump relationships by sharing parameters between token embeddings and position embeddings, but it lacks consideration for other dependencies. PalmTree models data dependencies by constructing Next Sequence Prediction (NSP) tasks on data flow graphs, but it sacrifices the ability to model complete assembly language context.
As for the masking strategy of Transformers~\cite{devlin2018bert} itself, a simple token-level mask can only model dependencies between individual tokens and cannot capture dependencies between instructions.
Consider the assembly snippet \texttt{test eax, eax; mov rcx, qword [rbp-num]; jne addr}. In cases like \texttt{test eax, [MASK]} or \texttt{jne [MASK]}, the model can accurately predict the masked tokens without contextual information. However, the model fails to learn the relationship between instructions \texttt{test eax, eax} and \texttt{jne addr}, which actually implies a conditional branch dependency.
In conclusion, as shown in Table \ref{tab:LiteratureReview}, There is still a lack of a comprehensive binary code embedding solution.


\begin{table}[]
\caption{Comparison of assembly language models.}
\label{tab:LiteratureReview}
\resizebox{.48\textwidth}{!}{
\begin{tabular}{c|cccc}
\toprule[1pt]
Method   & \begin{tabular}[c]{@{}c@{}}ISA\\ Knowledge\end{tabular} & \begin{tabular}[c]{@{}c@{}}Instruction\\ Boundary\end{tabular} & \begin{tabular}[c]{@{}c@{}}Implicit\\ Dependency\end{tabular} & \begin{tabular}[c]{@{}c@{}}Contextual\\ Inference\end{tabular} \\ \hline
word2vec & N                                                       & N                                                              & N                                                              & N                                                              \\
PalmTree & N                                                       & N                                                              & Partial                                                        & N                                                              \\
BinBert  & N                                                       & N                                                              & N                                                              & Y                                                              \\
jTrans   & N                                                       & Partial                                                        & Partial                                                        & Y                                                              \\
\sysname   & Y                                                       & Y                                                              & Y                                                              & Y                                                              \\ \bottomrule[1pt]
\end{tabular}
}
\end{table}

In this paper, we propose \sysname, a novel Transformer-based binary code embedding approach.
\sysname incorporates prior knowledge of assembly language into a Transformer model, and meanwhile is able to model the implicit dependencies between instructions with a novel pre-training task. 

\sysname incorporates explicit and implicit knowledge in two different ways respectively:
\begin{itemize}
    \item Explicit injection of token knowledge. This type of knowledge is generated based on the definition of the Instruction Set Architecture  (ISA), such as the instruction type and the relationship between registers \texttt{rax, eax, ax, al}, and is fed into the Transformer as additional inputs. By injecting this type of knowledge, the model gains the potential to learn token properties and instruction boundaries.
    \item Implicit injection of instruction knowledge. This type of knowledge is implicitly injected using instruction-level masks. By masking the entire instruction, the model gains an understanding of instruction boundaries and can model implicit relationships between instructions, such as conditional branch dependencies.
\end{itemize}




We conducted pre-training of \sysname on a large-scale binary code dataset and evaluated its performance 
on intrinsic tasks and real-world binary code analysis tasks.
The experimental results demonstrate that \sysname surpasses previous approaches in modeling assembly language with the lowest perplexity ($1 + 3.3e-4$) and produces high-quality binary code embeddings with the highest outlier detection accuracy ($86.6\%$). Through ablation experiments, we validated the effectiveness of the knowledge-aware design. 
Further, on 3 downstream tasks, i.e,  binary code similarity detection, function type recovery, and indirect call recognition, \sysname outperforms state-of-the-art solutions with 0.745 MRR@10000 ($\uparrow 5.2\%$), 0.876 accuracy ($\uparrow 6.8\%$) and 0.28 MRR@32 ($\uparrow 12.6\%$) respectively.

Overall, we demonstrate that injecting knowledge can effectively enhance the capabilities of assembly language models, providing a new direction for the development of language models for binary code. And with the advance of general-purpose large-scale models that offers new insights for domain models in the field of binary code embedding, we discuss three future research directions for binary code embedding: larger domain models, cost-effective models, and combination with general large language models.


In summary, this work makes the following contributions:
\begin{itemize}
    \item We propose a novel approach to incorporate prior knowledge into Transformers, able to model implicit dependencies in assembly language.
    It significantly improves the predictive capability of assembly language models, showing the promising prospects of knowledge fusion.
    \item We have conducted extensive experiments, and the results demonstrate that the embeddings generated by our pre-trained model outperform previous works and significantly enhance the performance of downstream tasks.
    \item We discuss several future research directions for binary code embedding.
    \item We open-source \href{https://github.com/Learner0x5a/kTrans-release}{\sysname} to facilitate future research.
\end{itemize}

\section{Background and Related Work}





\subsection{Binary Code Embedding}

Binary Code Embedding (BCE) is a technique that aims to map unstructured binary code into a low-dimensional space, where the binary code is represented as embeddings. 
BCE has numerous applications such as binary code similarity detection, type recovery, control-flow recovery and data-flow analysis, etc. 
Meanwhile, with the development of large language models, Transformer-based models~\cite{devlin2018bert} have emerged as popular methods for BCE for ability to learn complex features automatically.
This section will provide an overview of different methods for binary code embedding, including manual binary code embedding, non-Transformer-based learning-based binary code embedding, and Transformer-based learning-based binary code embedding.

\subsection{Manual Binary Code Embedding}

Manual binary code embedding involves representing binary code using manually constructed numerical features. Typically, these features are designed to capture syntactic and semantic information present in the binary code. Some of the commonly used features include:

\begin{itemize} 
\item Instruction N-grams: Instruction n-grams represent sequences of N consecutive instructions in the binary code. These features can capture local patterns and dependencies between instructions but may not be able to model long-range dependencies. 
\item Control flow graphs (CFGs): CFGs represent the control flow structure of the binary code as directed graphs, with basic blocks as nodes and control flow connections as edges between nodes. CFGs can capture the control flow structure of the binary code but may not model data dependencies and other semantic information. 
\item Function call graphs (FCGs): FCGs represent the calling relationships between functions in the binary code as directed graphs, with functions as nodes and function calls as edges. FCGs can capture high-level relationships between functions but may not model the detailed structure and semantics within individual functions. 
\end{itemize}

Despite their simplicity and ease of extraction, manual binary code embedding methods have several limitations. They require extensive domain expertise and are often task-specific, resulting in poor transferability between different binary code analysis tasks. Additionally, these approaches have limited adaptability and scalability for new challenges and advancements in the field of binary code analysis.

\subsection{Learning-based Binary Code Embedding}

This type of approaches automates the process of generating embeddings for binary code via representation learning, which learn features automatically from data, avoiding biases introduced by manual feature engineering. 

\paragraph{Non-Transformer-based}
This type of approaches can be divided into several categories based on the underlying learning algorithms they employ:

\begin{itemize} 
\item RNN-based: Methods like SAFE~\cite{massarelli2019safe} uses recurrent neural networks (RNNs) to model binary code sequences. RNN-based methods, such as LSTMs~\cite{LSTM} and GRUs~\cite{chung2014GRU}, can capture sequential dependencies in the data and are well-suited for modeling variable-length binary code sequences. However, RNNs may suffer from vanishing gradient issues when processing long sequences, limiting their ability to model long-range dependencies. 
\item CNN-based: Methods such as $\alpha $diff~\cite{liu2018alphadiff} use convolutional neural networks (CNNs) to model local patterns and relationships in the binary code. CNNs can capture spatial and hierarchical information using convolutional layers and pooling operations. However, CNNs may not be as effective in modeling sequential dependencies and long-range relationships in the binary code as RNNs or Transformers. 
\item GNN-based: Graph neural network (GNN) based methods, such as Gemini~\cite{xu2017neural} and VulSeeker~\cite{gao2018vulseeker}, model binary code using graph representations, such as control flow graphs (CFGs) and data flow graphs (DFGs). GNNs can capture complex relationships and dependencies between different parts of the binary code by processing graph-structured data. However, GNNs may require additional preprocessing steps to extract graph representations from binary code and can be computationally expensive for large graphs. \end{itemize}

\paragraph{Transformer-based}

Transformer-based binary code embedding methods leverage the powerful self-attention mechanism present in Transformers, which allows them to capture long-range dependencies and complex patterns in the binary code. Examples of Transformer-based methods include PalmTree~\cite{li2021palmtree}, jTrans~\cite{wang2022jtrans}, and COMBO~\cite{zhang2022combo}. These methods often employ pretraining-finetuning paradigms, which enable them to learn general-purpose binary code representations from large-scale, unlabeled data and then fine-tune these representations for specific binary code analysis tasks.

Some of the key advantages of Transformer-based methods over non-Transformer-based methods include:

\begin{itemize} 
\item Improved modeling of long-range dependencies: Transformers use self-attention mechanisms to model relationships between all pairs of tokens in the input sequence, allowing them to capture long-range dependencies more effectively than RNNs or CNNs. 
\item Scalability: Transformers can process input sequences in parallel, making them more computationally efficient and scalable compared to RNNs, which require sequential processing. 
\item Transfer learning: Transformer-based methods often employ transfer-learning paradigms such as pretraining-finetuning, which enable them to leverage large-scale, unlabeled data to learn general-purpose binary code representations and improve performance on specific binary code analysis tasks. 
\end{itemize}

However, Transformer-based binary code embedding methods also face some challenges:

\begin{itemize} 
\item Computational complexity: Although Transformers can process input sequences in parallel, their self-attention mechanism has quadratic complexity with respect to the sequence length, making them computationally expensive for long sequences. This can be mitigated to some extent using techniques like sparse attention or local attention. 
\item Prior knowledge incorporation: Most Transformer-based approaches treat binary code as natural language and directly apply language models to assembly language. This may overlook the prior knowledge inherent in binary code, such as opcode types, operand types, etc. 
\item Comprehending instructions: Many Transformer-based approaches lack a clear understanding of instruction boundaries, which may hinder their ability to model program execution behavior. Some methods, like PalmTree, require users to provide instruction boundaries, while others, like BinBert~\cite{artuso2022binbert}, lack information about instruction boundaries altogether. \end{itemize}


In summary, binary code embedding plays a crucial role in improving the performance of various binary code analysis tasks. While manual binary code embedding methods have certain limitations, learning-based binary code embedding methods, both non-Transformer-based and Transformer-based, have shown great promise in better capturing the structure and semantics of binary code. Continued research and development in this area can lead to more effective and comprehensive binary code embedding methods that can address the challenges and limitations of existing approaches.

In this work, to address these challenges and develop comprehensive embedding methods, we focus on incorporating prior knowledge of binary code, improving instruction understanding, and modeling implicit dependencies. 

\begin{figure*}[ht]
  \centering
  \includegraphics[width=\linewidth]{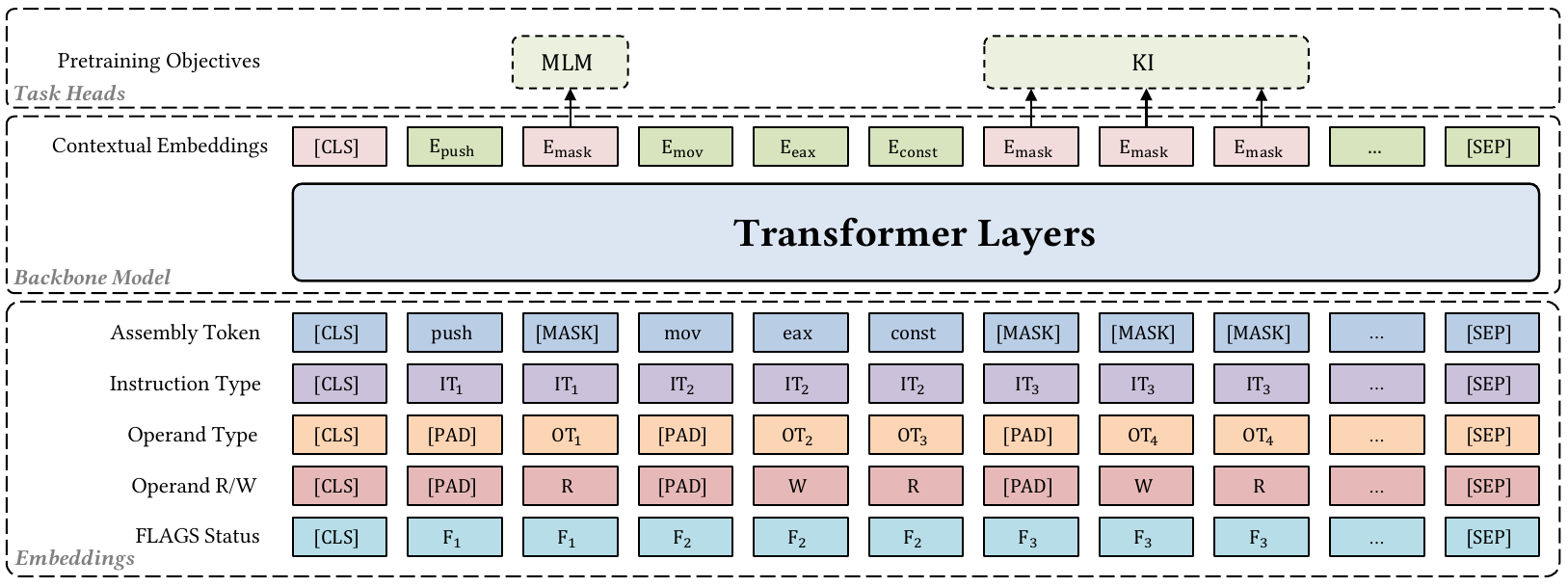}
  \caption{Overview of \sysname.}
  \label{fig:overview}
\end{figure*}

\subsection{Related Work}

\subsubsection{Representation Learning in NLP}

Representation learning in natural language processing (NLP) aims to learn continuous vector representations of words, phrases, or sentences that capture their semantic meaning. Techniques such as word2vec~\cite{mikolov2013word2vec}, GloVe~\cite{pennington2014glove}, and ELMo~\cite{peters-etal-2018-deep} have been widely used for various NLP tasks, including machine translation, sentiment analysis, and question-answering. These models are based on shallow neural networks or recurrent neural networks and rely on co-occurrence statistics to learn embeddings. More recently, Transformer-based models like BERT~\cite{devlin2018bert}, GPT~\cite{radfordimproving,radford2019language,NEURIPS2020_1457c0d6}, and RoBERTa~\cite{liu2019roberta} have achieved state-of-the-art results in numerous NLP benchmarks, showcasing the power of self-attention mechanisms in learning contextualized representations that can model long-range dependencies.

\subsubsection{Representation Learning for binary code}

Learning representations for binary code has been an active area of research in recent years. Early work focused on manual feature extraction methods~\cite{gao2008binhunt,xu2017neural,liu2018alphadiff,feng2016scalable,dullien2005graph}, where features like rawbytes, control flow graphs, the statistics of instructions, and function call graphs were used to capture the structure and semantics of the code. However, these manual approaches tend to be labor-intensive and may miss important semantic information.

Recent studies mostly focus on exploring learning-based techniques, including doc2vec~\cite{ding2019asm2vec}, CNN~\cite{liu2018alphadiff}, LSTM~\cite{massarelli2019safe,massarelli2019investigating}, graph neural networks~\cite{xu2017neural,gao2018vulseeker,marcelli2022cisco}, and Transformers~\cite{wang2022jtrans,zhang2022combo,artuso2022binbert,yu2020order,vulhawk}. These techniques leverage deep learning algorithms to learn embeddings, better capturing the structure and semantics of the binary code. The goal of representation learning for binary code is to enable improved performance in tasks like malware detection, reverse engineering, vulnerability analysis, and software similarity detection, etc.

\subsubsection{Representation Learning for source code}

Representation learning for source code aims to learn meaningful representations of source code that can be used in various software engineering tasks, such as code completion~\cite{svyatkovskiy2019pythia}, bug detection~\cite{Pradel2018DeepBugsAL}, program synthesis~\cite{austin2021program}, and code summarization~\cite{iyer-etal-2016-summarizing}. Techniques like code2vec~\cite{alon2019code2vec}, code2seq~\cite{alon2018code2seq}, and graph neural networks~\cite{Allamanis2017LearningTR} have been proposed to learn source code embeddings by capturing the syntactic and semantic information present in the code, such as abstract syntax trees, control flow graphs, and data flow graphs.

Recent work has also explored the use of Transformer-based models for source code representation learning~\cite{feng-etal-2020-codebert,Ahmad2021UnifiedPF}. These models demonstrate the potential of self-attention mechanisms in capturing complex patterns and long-range dependencies in source code, leading to improved performance on tasks like code summarization, code translation, and automated vulnerability detection. Additionally, the pretraining-finetuning paradigm employed by these Transformer-based models enables leveraging large-scale, unannotated source code data to improve generalization and robustness.

\subsubsection{Applications of Representation Learning to binary code analysis}
Representations of binary code can be used to solve various binary code analysis tasks.
PalmTree~\cite{li2021palmtree} proposes a Transformer-based general instruction embedding technique for multiple binary code ananlysis tasks.
Gemini~\cite{xu2017neural} expands the CFG with manually extracted features (e.g., number of instructions) to represent binary code,
and jTrans~\cite{wang2022jtrans} incorporates a unique jump-aware representation for control flow information in binary code.
ByteWeight \cite{shin2015boundary} and \cite{pei2020xda} identify function boundaries in binary code with RNNs and Transformers respectively,
and EKLAVYA~\cite{chua2017type} recovers function type signatures with word2vec to embed instructions.
CALLEE \cite{zhu2023callee} performs indirect call recognition with doc2vec to embed binary code slices,
and DeepVSA~\cite{guo2019deepvsa} facilitates value-set analysis with manual-designed instruction emebddings.
Except for the tasks mentioned before, binary code representations can also be applied to many other tasks such as vulnerability detection~\cite{wang2022jtrans,liu2018alphadiff,gao2018vulseeker, xu2017neural}, compiler provenance~\cite{he2022binprov,otsubo2020glassesx}, malware detection~\cite{xu2018deeprefiner,hu2009SMIT}, etc.

\section{Method}


\subsection{Overview}
To overcome the challenges mentioned in Section \ref{sec:intro}, we propose a new solution called \sysname to automatically incorporate knowledge in binary code.
\sysname is based on the Transformer-encoder architecture and follows a modular design to achieve high scalability.
As shown in Figure~\ref{fig:overview},
\sysname consists of three main modules: the embedding module, the backbone model, and the task head module.
The embedding module is responsible for the explicit injection of token-facts, the backbone model is responsible for generating contextual embeddings, and the task head module is responsible for the injection of implicit knowledge.

The embedding module generates three categories of input embeddings for the Transformer, and the final input embedding is obtained by their summation.
\begin{itemize}
    \item Assembly token embedding: For the textual sequence of assembly instructions, we employ a common tokenization method to obtain a token sequence and convert it into vectors, i.e., the token embedding.
    \item Explicit knowledge embedding: For the explicit knowledge contained in assembly instructions, such as the types of operation codes and operand types, we construct a vocabulary based on the definition of the ISA. We encode this knowledge as a sequence aligned with the assembly tokens to calculate the knowledge embeddings.
    \item Positional encoding: For positional encoding, we adopt the widely used positional encoding method from BERT. As for segment embedding, since our Transformer takes a single function as input and does not involve multiple sentences, segment embedding is not required.
\end{itemize}

The backbone model utilizes a Transformer model to integrate various embeddings and generates contextual embeddings. The Transformer model captures contextual information within the sequence using self-attention mechanism to generate contextual embeddings, which are then used for various downstream tasks.

The task head module is responsible for constructing pre-training tasks. During pre-training, \sysname consists of a masked language modeling (MLM) task and a knowledge injection (KI) task. 

\subsubsection{The embedding module}
\textbf{Assembly token embedding:} Following the common paradigm, we first normalize the assembly instructions to mitigate the out-of-vocabulary (OOV) problem. Then, we construct a vocabulary and convert the assembly instructions into tokens based on the vocabulary. We add special tokens to the token sequence of each assembly function, including [CLS] at the beginning and [SEP] at the end. Additional [PAD] tokens are appended to the end of the token sequence to ensure equal length for each sequence. Token sequences exceeding the maximum length limit are truncated. For tokens not found in the vocabulary, we uniformly represent them with the special [UNK] token. The final token sequence is then passed through a learnable embedding layer to obtain assembly token embeddings.

\begin{figure}[ht]
  \centering
  \includegraphics[width=\linewidth]{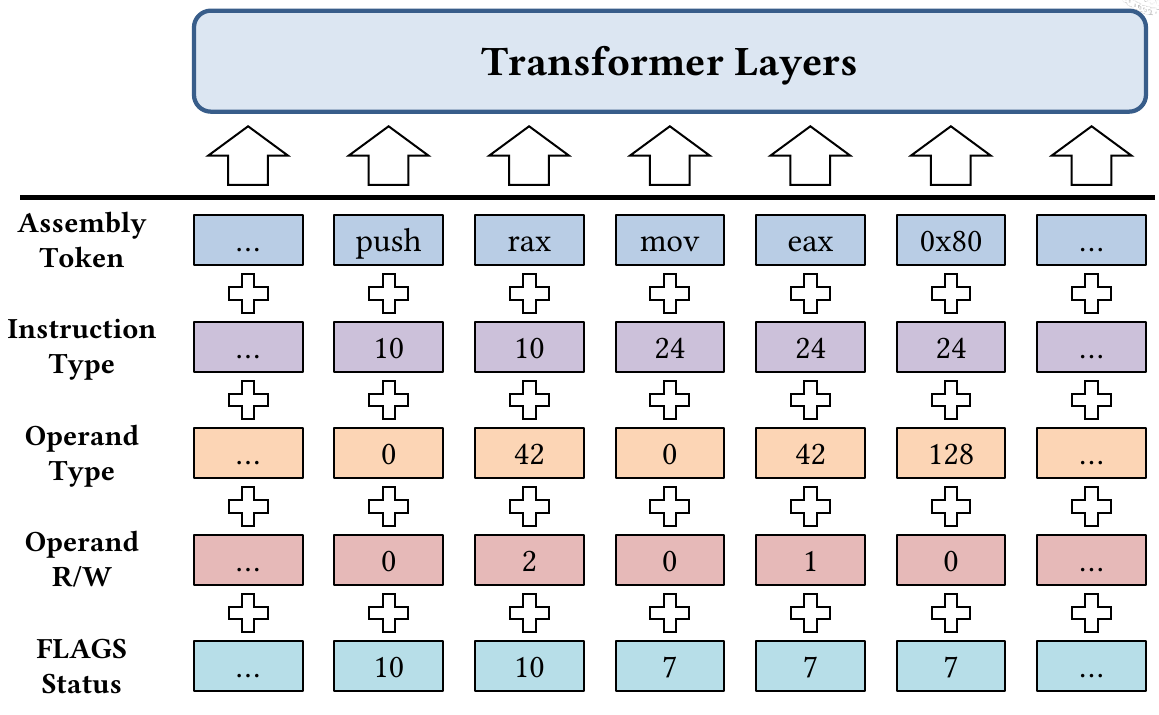}
  \caption{Explicit Knowledge Integration of \sysname.}
  \label{fig:ExplicitKnowledge}
\end{figure}

\textbf{Explicit knowledge embedding:} To leverage explicit prior knowledge to enhance the Transformer, we construct several knowledge sequences based on the ISA and perform similar embedding operations as the assembly token sequence. As shown in Figure~\ref{fig:ExplicitKnowledge}, we consider four types of explicit knowledge: opcode type, operand type, operand read/write status, and FLAGS register status. For each type of knowledge, we also build a vocabulary and add special tokens such as [CLS], [SEP], [PAD], [UNK], etc. After tokenization, the sequences are passed through their respective learnable embedding layers to generate knowledge embeddings.

The final embedding is obtained by summing up the embeddings from different sources: 
\[
E_{knowledge} = E_{InstructionType} + E_{OperandType} + E_{OperandRW} + E_{FLAGS}
\]
\[
E = E_{token} + E_{knowledge} + E_{position}
\]

\subsubsection{The backbone model}
We adopt a multi-layer Transformer encoder as the backbone model. The Transformer is a bidirectional language model based on the attention mechanism. With extensive training data and self-supervised learning, the Transformer can possess efficient knowledge transfer capabilities. The backbone model takes the final embedding $E$ as input and generates contextual embeddings through multiple layers of bidirectional Transformers.

\subsubsection{The task head module}



According to the research results of RoBERTa~\cite{liu2019roberta}, the Next Sentence Prediction (NSP) task is not an effective pre-training task. Therefore, we only use the Masked Language Modeling (MLM) task for pre-training and replace the NSP task with a Knowledge Integration (KI) task. The goal of the KI task is to enhance \sysname' understanding of instruction boundaries as well as implicit constraints between instructions.

\begin{figure*}[ht]
  \centering
  \includegraphics[width=.8\linewidth]{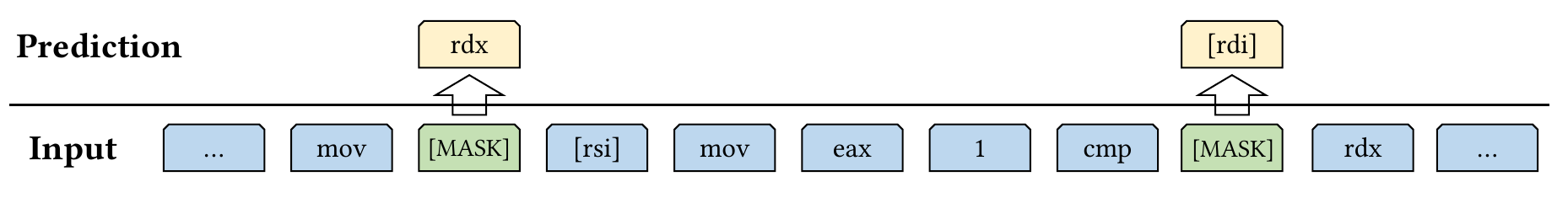}
  \caption{Masked Language Model (MLM).}
  \label{fig:MLM}
\end{figure*}

\textbf{MLM head}: We adopt the MLM task from BERT, where 15\% of tokens are randomly masked for the model to predict. Among these masked tokens, 80\% are replaced with the special token [MASK], 10\% are replaced with random tokens, and the remaining 10\% are unchanged. Let an assembly function be denoted as $\mathbf{f} = [x_{1},\cdots, x_{n}]$, where $x_i$ is the $i$-th token of $\mathbf{f}$, and $n$ is the number of tokens. We first select a random set of positions for $\mathbf{f}$ to mask out~(i.e., $\mathbf{m^{x}}$). 

\begin{equation}
\small
  \begin{aligned}
  \mathbf{f}^{\text{MLM}} = \text{REPLACE}(\mathbf{f},\mathbf{m^{x}},\texttt{[MASK]})\\
  \end{aligned}
\end{equation}
Based on these definitions, the MLM objective of reconstructing the masked tokens can be formulated as follows:
\begin{equation}
\small
  \min_{\theta} \mathcal{L}_{MLM}(\theta) = \sum_{i \in \mathbf{m_{x}}} -\log P(x_i|\mathbf{f}^{\text{MLM}})
\end{equation}
where $\mathbf{m^{x}}$ contains the indices of the masked tokens. 
An example of the masking process is presented in Figure~\ref{fig:MLM}.

\begin{figure*}[ht]
  \centering
  \includegraphics[width=.8\linewidth]{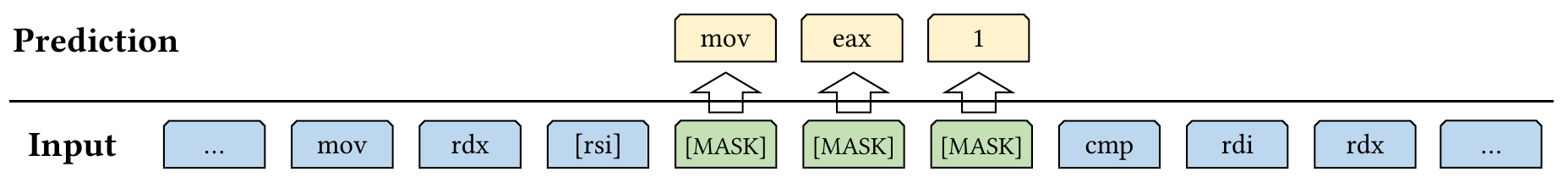}
  \caption{Knowledge Integration (KI).}
  \label{fig:KI}
\end{figure*}

\textbf{KI head}: Token-level masking cannot force the model to learn the concept of "instruction" and therefore cannot model the dependencies between instructions. Hence, we employ an instruction-level masking strategy. For a given assembly function, we randomly select 15\% of the instructions and replace all their tokens with a mask, while the tokens of the remaining instructions remain unchanged. Let an assembly function be denoted as $\mathbf{f} = [i_{1},\cdots, i_{n}]$, where $i_i$ is the $i$-th instruction of $\mathbf{f}$, and $n$ is the number of instructions. We first select a random set of positions for $\mathbf{f}$ to mask out~(i.e., $\mathbf{m^{i}}$). 

\begin{equation}
\small
  \begin{aligned}
  \mathbf{f}^{\text{KI}} = \text{REPLACE}(\mathbf{f},\mathbf{m^{i}},\texttt{[MASK]})\\
  \end{aligned}
\end{equation}
Based on these definitions, the KI objective of reconstructing the masked tokens can be formulated as follows:
\begin{equation}
\small
  \min_{\theta} \mathcal{L}_{KI}(\theta) = \sum_{j \in \mathbf{m_{i}}} -\log P(i_j|\mathbf{f}^{\text{KI}})
\end{equation}
where $\mathbf{m^{i}}$ contains the indices of the masked instructions. 
An example of the masking process is presented in Figure~\ref{fig:KI}.

The \textit{overall loss function} of \sysname in the pre-training phase is the summation of the MLM and KI objective functions:

\begin{equation}
  \min_{\theta} \mathcal{L}_{P}(\theta) = \mathcal{L}_{MLM}(\theta) + \mathcal{L}_{KI}(\theta)
\end{equation}

\subsection{Rationale}
\textbf{Explicit knowledge integration.}
The tokens and each type of explicit knowledge can be regarded as different modalities of the assembly, and data from a single modality can be incomplete or ambiguous. By integrating information from multiple modalities, multi-modal learning can reduce the impact of such issues. If we have sufficient data, multi-modal learning has been proven to be superior to single-modal learning~\cite{huang2021makes}.
Therefore, we expect explicit knowledge integration to generally enhance the model.

\textbf{Implicit knowledge integration.}
The objective of the MLM task is equivalent to maximizing the mutual information between the inputs and outputs of the Transformer encoder~\cite{kong2019mutual},
and different masking strategies (MLM and KI) actually provide different views of the assembly. 
Thus, optimizing the Transformer with multiple views, i.e. joint optimization, can generally result in larger mutual information than a single view.
In other words, if the generated embeddings can result in low losses for multiple views, it indicates that the representation capability of the Transformer is better, i.e., the representation capability of the generated embeddings is better.

\subsection{Dataset}
To pre-train \sysname, we use the BinaryCorp~\cite{wang2022jtrans} dataset to extract assembly code and knowledge.
The BinaryCorp dataset consists of over 10k projects and 26M functions of binary code, covering various types of targets such as editors, instant messengers, HTTP servers, web browsers, compilers, graphics libraries, crypto libraries, and so on.
We traverse the BinaryCorp dataset and extract functions from each binary program. For each function, we extracte the opcode types, operand types, operand read/write states, and EFLAGS register states of its instructions. During this process, functions that are too small (less than 5 instructions) will be filtered out.

\section{Implementation}
To achieve better scalability, \sysname does not rely on heavyweight program analysis.
It only requires the use of an arbitrary disassembly tool (e.g. IDA Pro~\cite{IDAPro}) to obtain function instructions and retrieve knowledge such as opcode types and operand types based on the ISA.


\textbf{Overcoming the OOV Problem.}
To overcome the Out-Of-Vocabulary (OOV) problem, after obtaining function boundaries in the binary program, we use Capstone~\cite{Capstone} to disassemble the instructions and extract explicit knowledge.
Based on the ISA, there are a finite number of states for opcode types, operand types, operand read/write status, and EFLAGS register status. Therefore, we only need to normalize the assembly instruction text. We adopt three strategies for normalization: (1) treat mnemonics and operands as tokens, (2) replace all constants with "const", and (3) remove commas between operands.

In summary, the vocabulary size (including special tokens) of assembly tokens is 15,511, and the vocabulary sizes (including special tokens) of various types of knowledge are shown in the Table~\ref{tab:vocab}.

\begin{table}[ht]
\caption{Vocabulary Sizes.}
\label{tab:vocab}
\centering
\begin{tabular}{c|c}
\toprule[1pt]
Vocab            & Size   \\ \hline
Assembly Token   & 15,511 \\
Instruction Type & 1,002   \\
Operand Type     & 12     \\
Operand R/W      & 9      \\
FLAGS Status     & 56     \\ \bottomrule[1pt]
\end{tabular}
\end{table}

\section{Evaluation}
\subsection{Evaluation Methodology}
In the evaluation, we address the following four research questions:
\begin{itemize}
    \item \textbf{RQ1:} How does \sysname perform in modeling assembly language? (see \ref{eval:RQ1})
    \item \textbf{RQ2:} What is the Impact of the Knowledge-Aware Design? (see \ref{eval:RQ2})
    \item \textbf{RQ3:} How is the quality of the embeddings generated by \sysname? (see \ref{eval:RQ3})
    \item \textbf{RQ4:} Can \sysname effectively improve the performance of downstream tasks? (see \ref{eval:RQ4})
\end{itemize}
To answer \textbf{RQ1}, we calculate the perplexity of \sysname and compare it with several baselines. To answer \textbf{RQ2}, we conduct ablation experiments to evaluate the effects of explicit knowledge integration and implicit knowledge integration separately. To answer \textbf{RQ3}, we perform quantitative evaluation using outlier detection and qualitative evaluation using t-SNE visualization. Finally, to answer \textbf{RQ4}, we apply \sysname to three types of binary code analysis tasks: binary code similarity identification, function type recovery, and indirect call identification. We evaluate the performance of zero-shot learning (directly using the embeddings generated by \sysname) as well as the performance after fine-tuning \sysname.

\subsection{Evaluation Setup}
\subsubsection{Dataset}
To pre-train \sysname, we choose the BinaryCorp dataset, which consists of a large collection of binary programs obtained through automated compilation. The dataset contains a total of 48,130 binary programs, approximately 26 million binary functions, covering two compilers (gcc and clang) and five compilation optimization levels (O0-O3, Os). The detailed statistics for the training and testing sets are provided in Table~\ref{tab:BinaryCorp_stat}.

\begin{table}[h]
\centering
\small
  \caption{Statistics on the number of projects, binaries and functions of the datasets.}
  \begin{tabular}{c|ccccc}
  \toprule[1pt]
  Datasets              & \# Projects & \# Binaries & \# Functions  \\ \hline
  BinaryCorp~Train & 7,845       & 38,455      & 21,085,338    \\
  BinaryCorp~Test  & 1,974       & 9,675       & 4,791,673       \\ 
  \bottomrule[1pt]
  \end{tabular}
  \label{tab:BinaryCorp_stat}
\end{table}

Compared to datasets such as BinKit~\cite{kim:tse:2022} and Cisco~\cite{marcelli2022cisco}, BinaryCorp covers a significantly larger number of projects, making it a more diverse and realistic dataset. As a result, it is better suited for evaluating the scalability and efficacy of binary code language models.

\subsubsection{Baselines in Modeling Assembly}

In terms of modeling assembly, we compare \sysname with two representative baselines:
\begin{itemize}
    \item Non-Transformer-based baseline: word2vec~\cite{mikolov2013word2vec}. By treating mnemonics and operands as tokens, we can obtain their embeddings using word2vec. The assembly instruction embedding and the embedding of the instruction sequence can be obtained by taking the average of their token embeddings respectively.
    \item Transformer-based baseline: PalmTree~\cite{li2021palmtree}. PalmTree provides instruction-level embeddings using a Transformer. To ensure fairness, we retrained the PalmTree model on the BinaryCorp dataset. Similarly, the embedding of the instruction sequence can be obtained by taking the average of the instruction embeddings.
\end{itemize}

\subsubsection{System Configuration}
\sysname consists of 12 Transformer layers with a hidden dimension of 768. The number of heads in the multi-head attention mechanism is set to 8. During the pre-training process, the warm-up ratio is 0.01, the initial learning rate is 3e-5, and the learning rate decay is 0.01.

All experiments are conducted on a Linux server running Ubuntu 18.04 with Intel(R) Xeon(R) Gold 6154 CPU @ 3.00GHz, 512GB RAM, and 4 NVIDIA V100 GPUs.

\subsection{\textbf{RQ1:} How does \sysname perform in modeling assembly language?} \label{eval:RQ1}
We use perplexity to evaluate the performance of each model in modeling assembly. All models are trained for 3 epochs on the training set of binary corp and then perplexity is calculated on the test set.

Perplexity is defined as the exponentiation of the average logarithmic probability per symbol of a test set, normalized by the number of symbols: 
\[
PPL = 2^{-\frac{1}{N} \sum_i \log_2(P(x_i))}
\]
where N is the total number of tokens in the test set, and $P(x_i)$ represents the probability assigned by the language model to the token $x_i$. 
Perplexity measures how surprised a language model is when it encounters new tokens. A lower perplexity indicates that the model is better at predicting the next token, while a higher perplexity suggests that the model is more uncertain and less accurate in its predictions.
Theoretically, the lowest possible value of perplexity is 1, and the closer the perplexity value is to 1, the better the model's performance in modeling assembly.

As shown in Table \ref{tab:ppl}, word2vec has the highest perplexity, followed by PalmTree with slightly lower perplexity, and \sysname has the lowest perplexity, $1+3.30e^{-4}$. Therefore, \sysname performs the best in modeling assembly.

\begin{table}[]
\centering
\small
\caption{Perplexity Evaluation.}
\label{tab:ppl}
\renewcommand\arraystretch{1.25}
\begin{tabular}{c|c}
\toprule[1pt]
Model    & $PPL-1$    \\ \hline
word2vec      & $9.01e^{-3}$ \\
PalmTree & $9.05e^{-3}$ \\
\sysname   & $3.30e^{-4}$ \\ \bottomrule[1pt]
\end{tabular}
\end{table}

\subsection{\textbf{RQ2:} What is the Impact of the Knowledge-Aware Design?}\label{eval:RQ2}
To evaluate the contribution of knowledge injection to the model's capability, we conduct a set of ablation experiments and evaluate the model's performance of loss on the test set. Considering computational costs, the hidden layer dimension of the models in the ablation experiments is set to 128 to speed up training. We compare the loss of four different settings: 
(A) Original Transformer;
(B) Transformer with explicit knowledge;
(C) Transformer with implicit knowledge; 
(D) Transformer with explicit and implicit knowledge.

By comparing (A, B) and (C, D), we can assess the impact of explicit knowledge. Similarly, by comparing (A, C) and (B, D), we can evaluate the effect of implicit knowledge.

\begin{figure}[t]
  \centering
  \includegraphics[width=\linewidth]{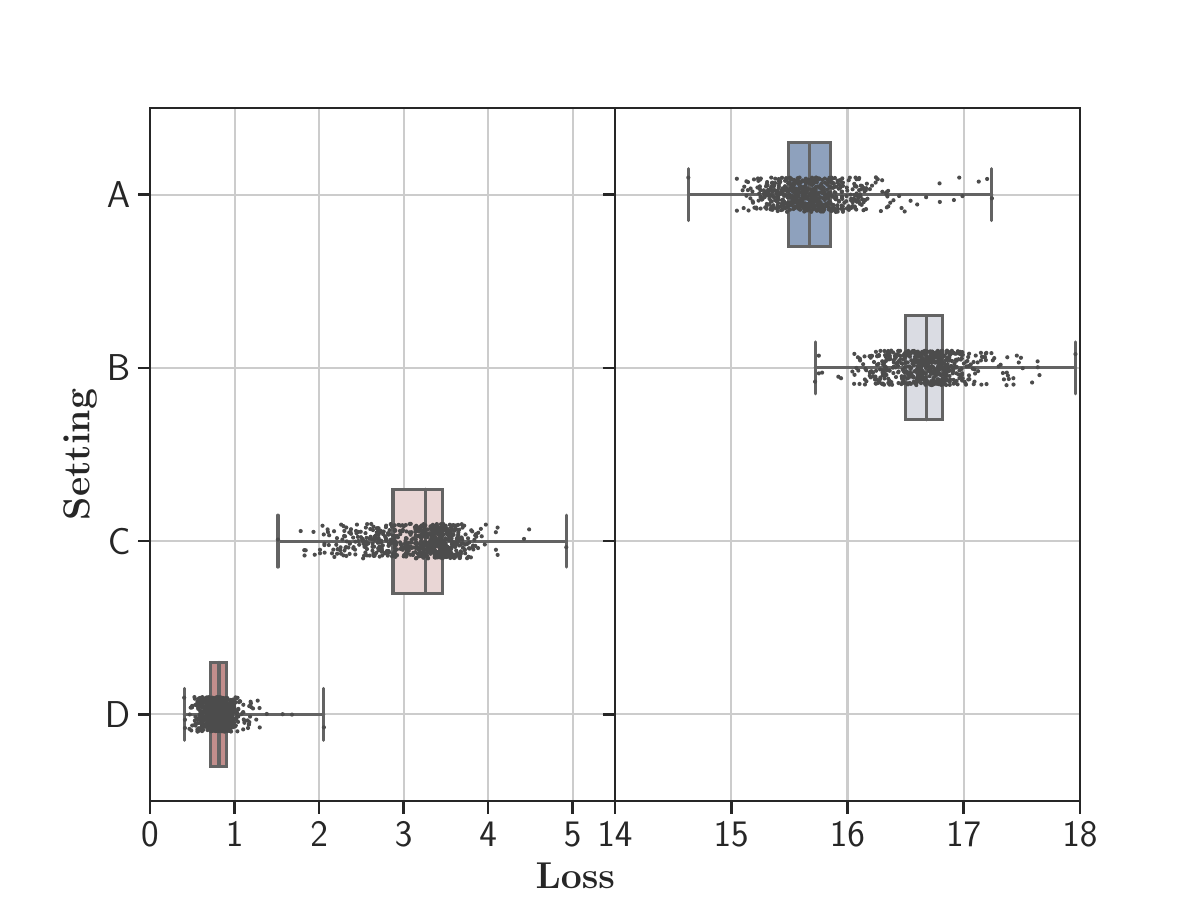}
  \caption{The distributions of losses on the test set under different settings.}
  \label{fig:ablation}
\end{figure}

As shown in Figure \ref{fig:ablation} and Table~\ref{tab:ablation}, for (A, C) and (B, D), introducing implicit knowledge results in a significant decrease in the model's loss, indicating that incorporating implicit knowledge through the KI task can significantly enhance the model's capability. In the case of comparison (A, B), adding explicit knowledge leads to a slight increase in the loss, possibly because introducing additional knowledge in the MLM task alone could cause overfitting. However, for comparison (C, D), after introducing the KI task, explicit knowledge significantly reduces the model's loss, indicating that the combination of explicit knowledge and the KI task can greatly enhance the model's capability. 


\begin{table}[h]
\caption{Average losses on the test set under different settings.}
\label{tab:ablation}
\centering
\small
\begin{tabular}{c|cc|c}
\toprule[1pt]
Setting & \begin{tabular}[c]{@{}c@{}}Explict\\ Knowledge\end{tabular} & \begin{tabular}[c]{@{}c@{}}Implicit\\ Knowledge\end{tabular} & \begin{tabular}[c]{@{}c@{}}Evaluation \\ Loss\end{tabular} \\ \hline
A       & \xmark                 & \xmark                  & 15.8      \\
B       & \cmark                 & \xmark                  & 16.7      \\
C       & \xmark                 & \cmark                   & 3.11      \\
D       & \cmark                  & \cmark                   & 0.858     \\ \bottomrule[1pt]
\end{tabular}
\end{table}

Overall, these results show that incorporating external knowledge into the assembly language sequence modeling is highly beneficial for our model.

\subsection{\textbf{RQ3:} How is the quality of the embeddings generated by \sysname?} \label{eval:RQ3}
\subsubsection{Outlier Detection}

Outlier detection is a quantitative evaluation method for assessing the quality of embeddings.
Formally, given a set of tokens $T = {t_1, t_2, \ldots, t_n, t_{n+1}}$, 
assume that $t_1, \ldots, t_n$ belong to the same cluster, and $t_{n+1}$ is the outlier,
the task aims to identify the outlier that does not belong to the same group as the remaining tokens. 

The performance of the outlier detection task can be evaluated using accuracy. Ideally, if outliers in all the groups are correctly identified, the accuracy should be 1.
Following the setup in PalmTree, we evaluate both opcode outlier detection and operand outlier detection.

We construct an instruction set from the binaries in the test set and sample 50,000 groups of data. For opcode outlier detection, each group of data consists of four instructions with the same opcode and one instruction with a different opcode. For operand outlier detection, each group of data consists of four instructions with the same operand type and one instruction with a different operand type.

\begin{figure*}
     \centering
     \begin{subfigure}[b]{0.45\textwidth}
         \centering
         \includegraphics[width=\textwidth]{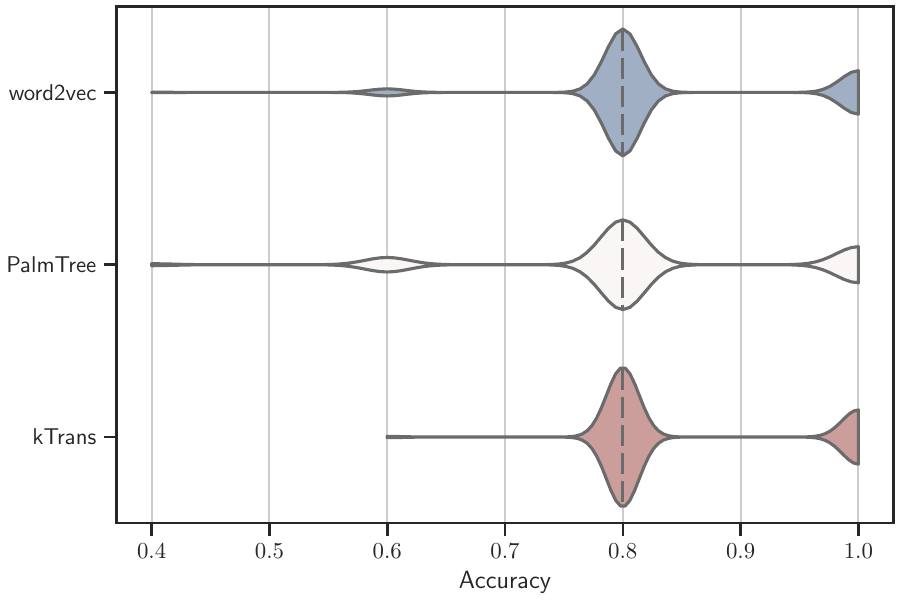}
         \caption{Opcode}
         \label{fig:OutlierOpcode}
     \end{subfigure}
     \hfill
     \begin{subfigure}[b]{0.45\textwidth}
         \centering
         \includegraphics[width=\textwidth]{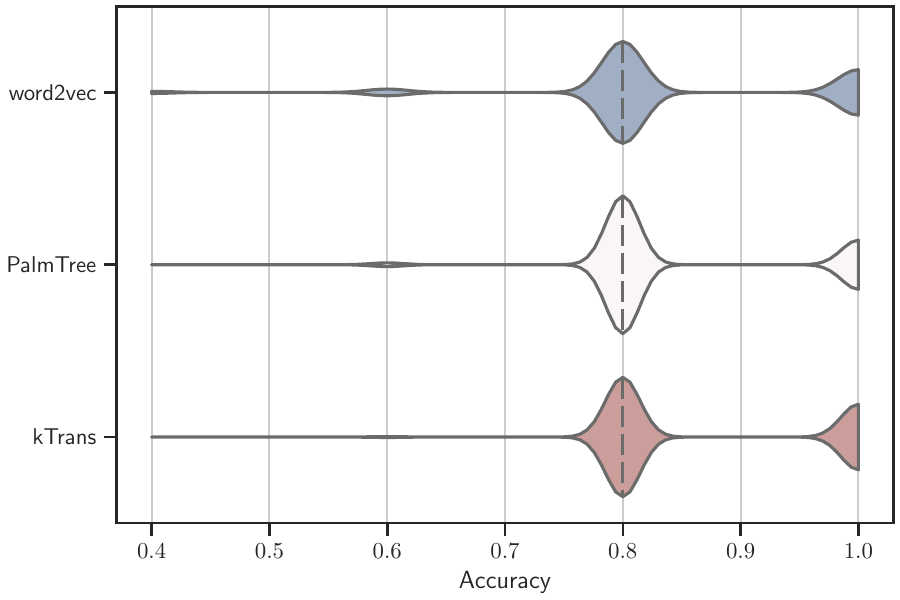}
         \caption{Operand}
         \label{fig:OutlierOpnd}
     \end{subfigure}
     \hfill
     \caption{Accuracy of outlier detection.}
     \label{fig:outlier}
\end{figure*}

Table \ref{tab:outlier} shows the average accuracy of outlier detection for each model. Figure~\ref{fig:outlier} displays the distribution of outlier detection accuracy for each model.
As shown, \sysname still achieves the best performance, with an accuracy of 86.1\% for opcode outlier detection and 86.56\% for operand outlier detection. PalmTree performs even worse than word2vec on this task.

\begin{table}[ht]
\centering
\caption{Average accuracy of different models for outlier detection. Acc-Opcode denotes the average accuracy of opcode outlier detection, and Acc-Opnd denotes the average accuracy of operand outlier detection.}
\label{tab:outlier}
\small
\begin{tabular}{c|cc}
\toprule[1pt]
Model    & Acc-Opcode & Acc-Opnd \\ \hline
word2vec      & 0.840     & 0.846   \\
PalmTree & 0.829     & 0.847   \\
\sysname   & 0.861     & 0.866   \\ \bottomrule[1pt]
\end{tabular}
\end{table}

\begin{figure*}[t]
     \centering
     \begin{subfigure}[b]{0.3\textwidth}
         \centering
         \includegraphics[width=\textwidth]{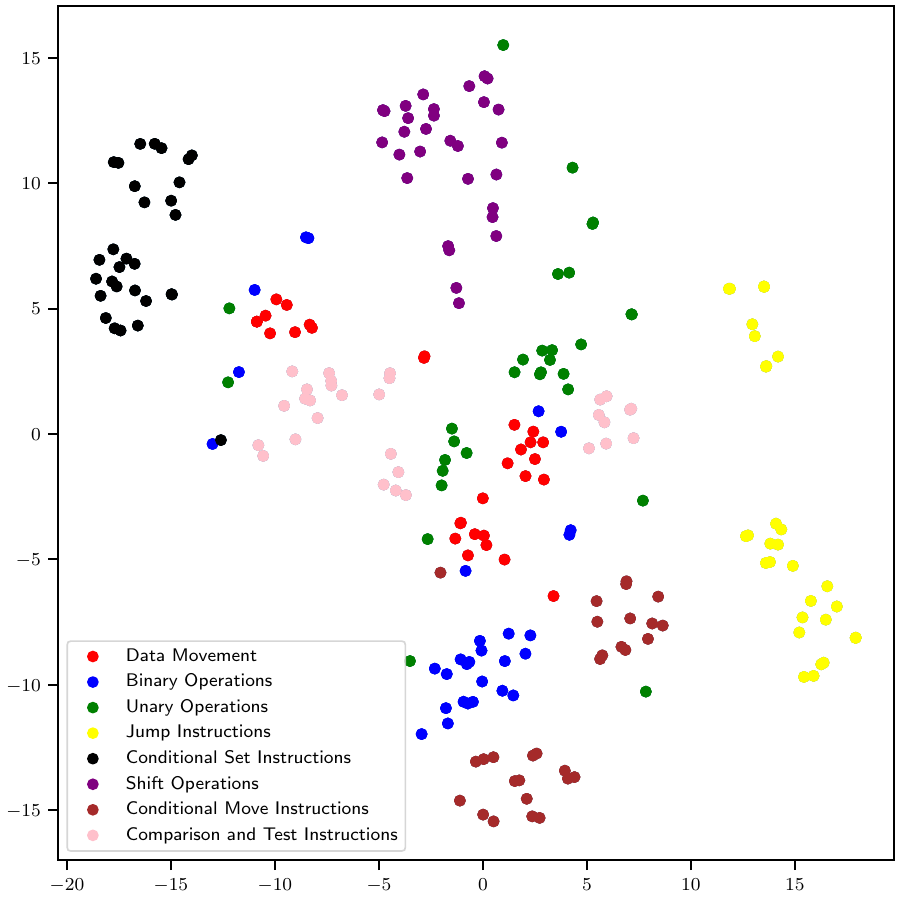}
         \caption{word2vec}
         \label{fig:tsne:w2v}
     \end{subfigure}
     \hfill
     \begin{subfigure}[b]{0.3\textwidth}
         \centering
         \includegraphics[width=\textwidth]{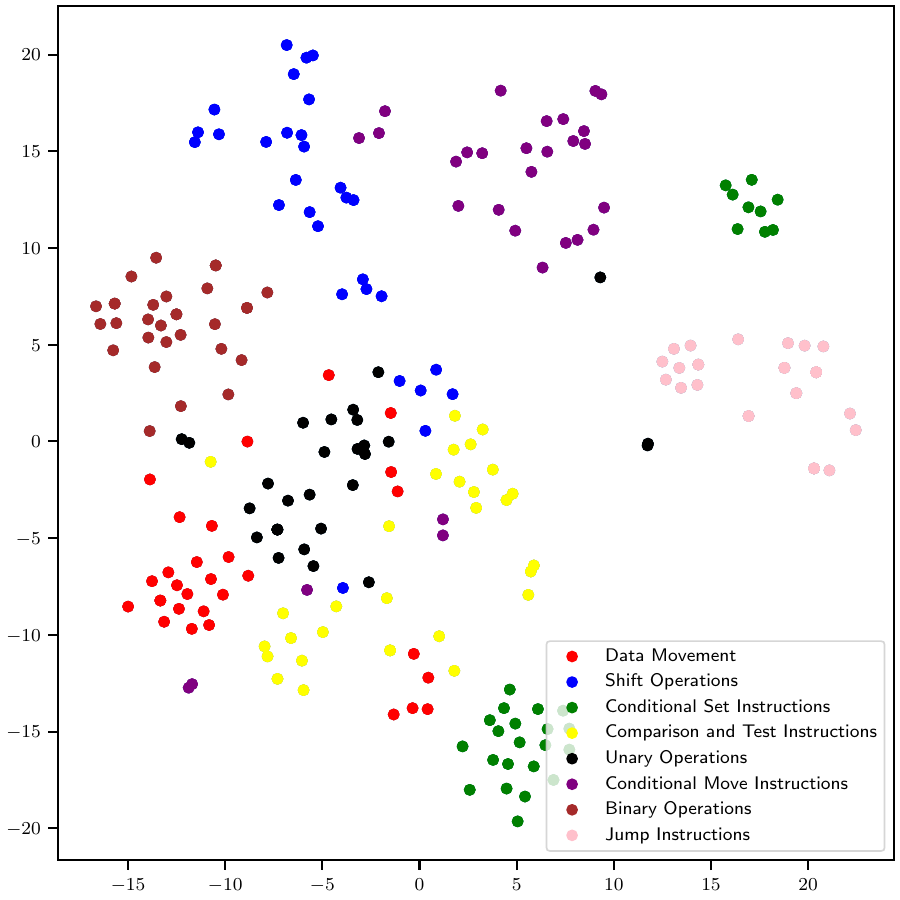}
         \caption{PalmTree}
         \label{fig:tsne:palmtree}
     \end{subfigure}
     \hfill
     \begin{subfigure}[b]{0.3\textwidth}
         \centering
         \includegraphics[width=\textwidth]{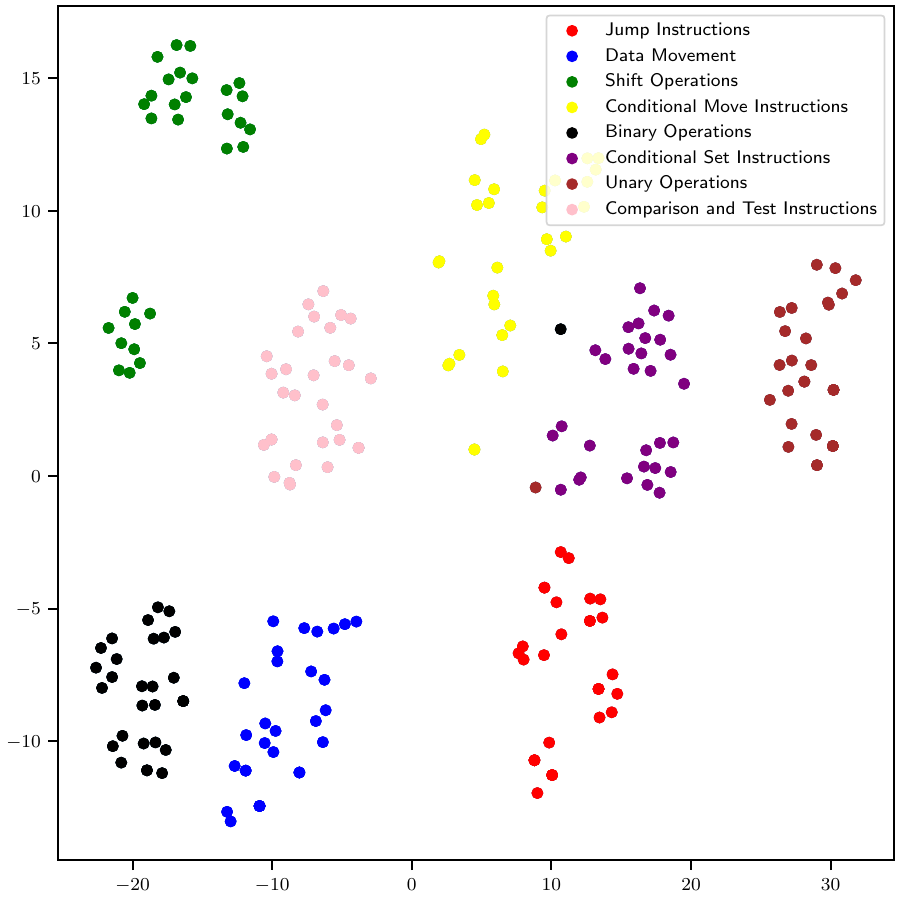}
         \caption{\sysname}
         \label{fig:tsne:ktrans}
     \end{subfigure}
     \hfill
     \caption{t-SNE visualization of embeddings.}
     \label{fig:tsne}
\end{figure*}

\subsubsection{Qualitative Analysis through t-SNE visualization}


We also observe the distribution of instruction embeddings and analyze the quality of embeddings using visualization.

t-distributed stochastic neighbor embedding (t-SNE) is a statistical method that visualizes high-dimensional data by assigning each data point a location on a two- or three-dimensional map.
We construct an instruction set from binaries in the test set and sample 30 instances from each of the eight common instruction classes. Then we use t-SNE to observe the distribution of their embeddings.

As shown in Figure \ref{fig:tsne}, compared to word2vec and PalmTree, the embeddings generated by \sysname exhibits clearer class boundaries and denser distributions within the same class. While PalmTree fails to distinguish between "Data Movement", "Comparison and Test Instructions" and "Unary Operations", and word2vec mixes "Data Movement" and "Conditional Movement Instructions" together.

Through both quantitative and qualitative analyses, we can conclude that \sysname can produce high-quality binary code embeddings that surpass previous work.

\subsection{\textbf{RQ4:} Can \sysname effectively improve the performance of downstream tasks?} \label{eval:RQ4}
\subsubsection{Binary Code Similarity Detection}

\begin{table*}[h]
\centering
\caption{Results of different binary similarity detection methods on BinaryCorp (Poolsize=32).}
\label{tab:binsim32}
\scalebox{0.8}{
\begin{tabular}{c|ccccccc|ccccccc}
\toprule[1pt]
\multirow{2}{*}{Model} & \multicolumn{7}{c|}{MRR}                                                                                                                  & \multicolumn{7}{c}{Recall@1}                                                                                                                  \\ \cline{2-15} 
                       & O0,O3          & O1,O3          & O2,O3          & O0,Os          & O1,Os          & \multicolumn{1}{c|}{O2,Os}          & Average        & O0,O3          & O1,O3          & O2,O3          & O0,Os          & O1,Os          & \multicolumn{1}{c|}{O2,Os}          & Average        \\ \hline
Gemini                 & 0.402          & 0.643          & 0.835          & 0.469          & 0.564          & \multicolumn{1}{c|}{0.628}          & 0.590           & 0.263          & 0.528          & 0.768          & 0.322          & 0.441          & \multicolumn{1}{c|}{0.518}          & 0.473          \\
word2vec                    & 0.433          & 0.841          & 0.952          & 0.474          & 0.778          & \multicolumn{1}{c|}{0.849}          & 0.721          & 0.315          & 0.789          & 0.936          & 0.352          & 0.716          & \multicolumn{1}{c|}{0.798}          & 0.651          \\
jTrans-zero            & 0.594          & 0.841          & 0.962          & 0.649          & 0.850           & \multicolumn{1}{c|}{0.891}          & 0.797          & 0.499          & 0.803          & 0.945          & 0.566          & 0.808          & \multicolumn{1}{c|}{0.853}          & 0.746          \\
\sysname-zero            & 0.537          & 0.951          & 0.981          & 0.570           & 0.908          & \multicolumn{1}{c|}{0.910}           & 0.810           & 0.402          & 0.932          & 0.973          & 0.435          & 0.877          & \multicolumn{1}{c|}{0.881}          & 0.750           \\
PalmTree               & 0.688          & 0.873          & 0.956          & 0.729          & 0.850           & \multicolumn{1}{c|}{0.904}          & 0.833          & 0.567          & 0.813          & 0.936          & 0.618          & 0.777          & \multicolumn{1}{c|}{0.859}          & 0.762          \\
jTrans                 & 0.947          & 0.976          & 0.985          & 0.956          & 0.979          & \multicolumn{1}{c|}{0.977}          & 0.970           & 0.913          & 0.960           & 0.974          & 0.927          & 0.964          & \multicolumn{1}{c|}{0.961}          & 0.949          \\
\sysname                 & \textbf{0.975} & \textbf{0.991} & \textbf{0.996} & \textbf{0.984} & \textbf{0.992} & \multicolumn{1}{c|}{\textbf{0.991}} & \textbf{0.988} & \textbf{0.956} & \textbf{0.984} & \textbf{0.993} & \textbf{0.971} & \textbf{0.985} & \multicolumn{1}{c|}{\textbf{0.984}} & \textbf{0.979} \\ \bottomrule[1pt]
\end{tabular}
}
\end{table*}

\begin{table*}[h]
\centering
\caption{Results of different binary similarity detection methods on BinaryCorp (Poolsize=10,000).}
\label{tab:binsim10000}
\scalebox{0.8}{
\begin{tabular}{c|ccccccc|ccccccc}
\toprule[1pt]
\multirow{2}{*}{Model} & \multicolumn{7}{c|}{MRR}                                                                                                               & \multicolumn{7}{c}{Recall@1}                                                                                                                 \\ \cline{2-15} 
                       & O0,O3          & O1,O3        & O2,O3         & O0,Os          & O1,Os          & \multicolumn{1}{c|}{O2,Os}          & Average        & O0,O3          & O1,O3          & O2,O3         & O0,Os          & O1,Os          & \multicolumn{1}{c|}{O2,Os}          & Average        \\ \hline
Gemini                 & 0.072          & 0.189        & 0.474         & 0.069          & 0.147          & \multicolumn{1}{c|}{0.202}          & 0.192          & 0.058          & 0.148          & 0.420          & 0.051          & 0.115          & \multicolumn{1}{c|}{0.162}          & 0.159          \\
word2vec                    & 0.119          & 0.423        & 0.693         & 0.097          & 0.369          & \multicolumn{1}{c|}{0.446}          & 0.358          & 0.099          & 0.367          & 0.628         & 0.086          & 0.307          & \multicolumn{1}{c|}{0.403}          & 0.315          \\
jTrans-zero            & 0.215          & 0.570         & 0.759         & 0.233          & 0.571          & \multicolumn{1}{c|}{0.563}          & 0.485          & 0.167          & 0.503          & 0.701         & 0.175          & 0.507          & \multicolumn{1}{c|}{0.500}            & 0.426          \\
\sysname-zero            & 0.129          & 0.702        & 0.835         & 0.111          & 0.645          & \multicolumn{1}{c|}{0.664}          & 0.514          & 0.105          & 0.633          & 0.789         & 0.090           & 0.538          & \multicolumn{1}{c|}{0.610}           & 0.461          \\
PalmTree               & 0.130           & 0.403        & 0.677         & 0.152          & 0.355          & \multicolumn{1}{c|}{0.496}          & 0.369          & 0.083          & 0.326          & 0.609         & 0.097          & 0.281          & \multicolumn{1}{c|}{0.420}           & 0.303          \\
jTrans                 & 0.584          & 0.734        & 0.792         & 0.627          & 0.709          & \multicolumn{1}{c|}{0.710}           & 0.693          & 0.499          & 0.668          & 0.736         & 0.550           & 0.648          & \multicolumn{1}{c|}{0.648}          & 0.625          \\
\sysname                 & \textbf{0.588} & \textbf{0.800} & \textbf{0.870} & \textbf{0.657} & \textbf{0.777} & \multicolumn{1}{c|}{\textbf{0.777}} & \textbf{0.745} & \textbf{0.493} & \textbf{0.732} & \textbf{0.820} & \textbf{0.565} & \textbf{0.712} & \multicolumn{1}{c|}{\textbf{0.716}} & \textbf{0.673} \\ \bottomrule[1pt]
\end{tabular}
}
\end{table*}

\begin{figure*}[]
  \centering
  \includegraphics[width=\linewidth]{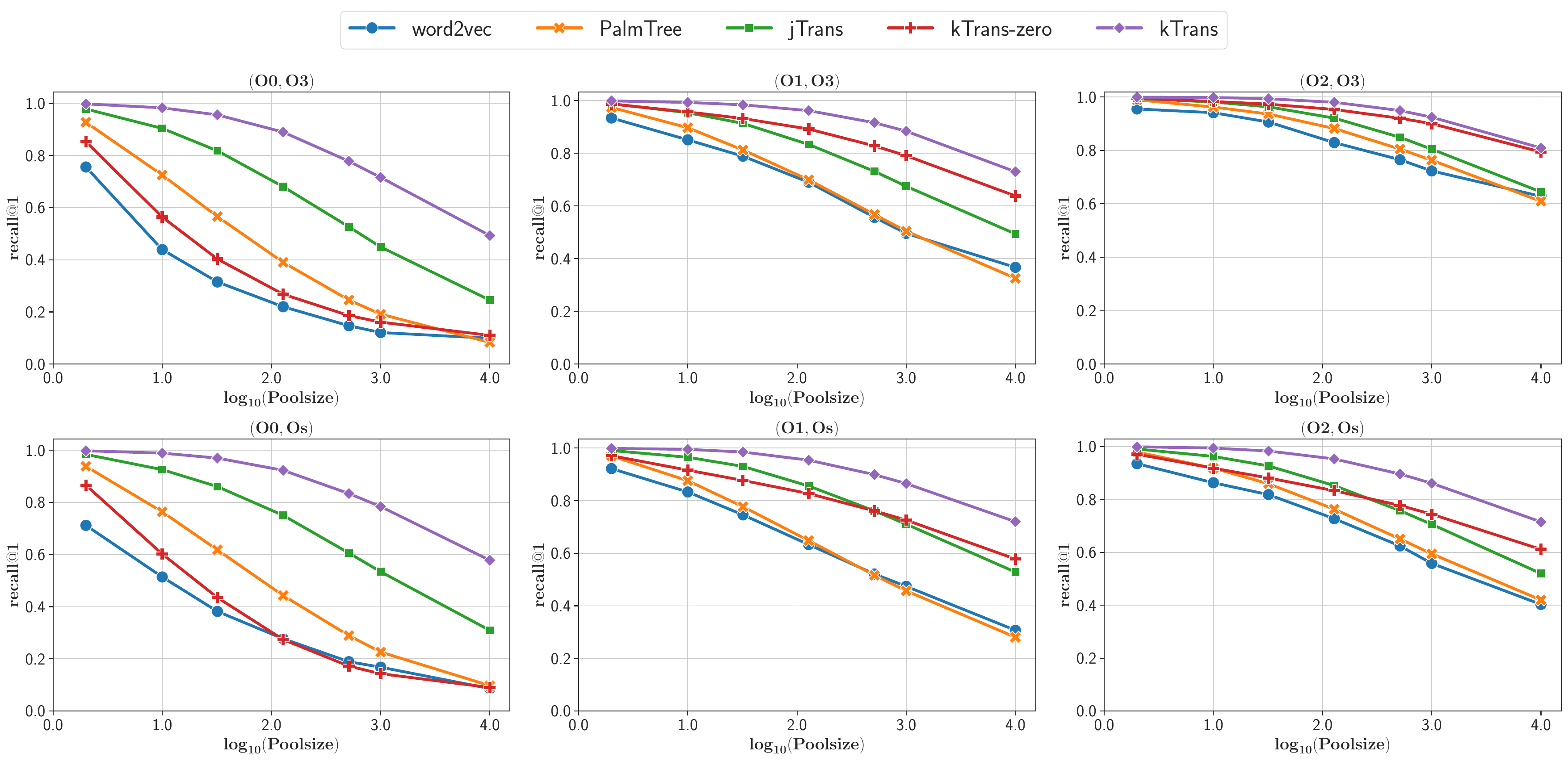}
  \caption{The performance of different models for binary code similarity detection with respect to pool size.}
  \label{fig:binsim}
\end{figure*}
Binary code similarity detection (BCSD) aims to determine the level of similarity between two binary code snippets.

Based on BinaryCorp, we evaluate the performance of \sysname on BCSD. In addition to word2vec and PalmTree, we compare \sysname with a GNN-based approach Gemini and the state-of-the-art BCSD solution jTrans. The evaluation metrics used are MRR (Mean Reciprocal Rank) and Recall@k, defined as follows: 
\[
\text{Recall}@k = \frac{1}{|\mathcal{F}|} \sum_{f_{i}\in \mathcal{F}}\mathbb{I}(\text{Rank}_{f^{gt}_{i}} \le k)
\]
\[
\text{MRR} = \frac{1}{|\mathcal{F}|} \sum_{f_{i} \in \mathcal{F}} \frac{1}{Rank_{f^{gt}_{i}}}
\]
where $\mathcal{F}$ is a binary function pool, $\mathcal{G}$ is the ground truth binary function pool,
$\mathbb{I}$ is the indicator function.
And we denote a query function as $f_{i} \in \mathcal{F}$ and its corresponding ground truth function as $ f^{gt}_{i} \in \mathcal{G}$. 

As shown in Table \ref{tab:binsim32}, with a pool size of 32, \sysname achieves an average MRR of 0.988 and an average Recall@1 of 0.979, surpassing all other models. In the zero-shot scenario, except for (O0, O3), \sysname outperforms zero-shot jTrans in all other settings. PalmTree performs better than word2vec and Gemini but only outperforms zero-shot \sysname by 0.01-0.02, far behind jTrans and \sysname.
When pool size comes to 10,000 (Table~\ref{tab:binsim10000}), \sysname still has the best performance among all methods with MRR of 0.745 and Recall@1 of 0.673, 
while PalmTree's MRR drops sharply to 0.369, making it unpractical in the real-world BCSD scenario.

We also study the relationship between pool size and performance of different BCSD methods.
As shown in Figure~\ref{fig:binsim}, \sysname achieves best performance under all settings,
and even \sysname-zero can outperform jTrans in (O1,O3) and (O2,O3). When pool size comes to 10,000,
\sysname-zero can outperform jTrans in (O1,Os) and (O2,Os) as well.

Therefore, these results demonstrate that incorporating external knowledge into the embedding can enhance the performance of the BCSD task.

\subsubsection{Function Type Recovery}

Function type recovery (FTR) aims to determine the number and types of arguments for a function in binary code.

We compare \sysname, PalmTree, and EKLAVYA in this task. EKLAVYA utilizes word2vec as the instruction embedding method and predicts function types using a Recurrent Neural Network (RNN). Hence, we replace the instruction embedding method with \sysname and PalmTree for comparison.

The dataset used in this experiment is provided by EKLAVYA and consists of 8 projects and 2,312 binaries. In the papers of EKLAVYA and PalmTree, they employ binary split to divide the dataset into training and testing sets, which may result in overlap between the two sets due to shared code among different binaries within the same project. Therefore, we perform a new split based on projects, selecting 6 projects randomly as the training set and the remaining 2 projects as the testing set. The evaluation metric used is the accuracy of function type prediction.

As shown in Table \ref{tab:functype}, \sysname achieves the highest accuracy on the testing set, at 87.6\%. PalmTree and EKLAVYA (word2vec) achieve similar performance at 80\%, but still fall short compared to zero-shot \sysname. Additionally, by comparing the accuracy between the training and testing sets, we observe that PalmTree and EKLAVYA suffer from significant overfitting.

Therefore, these results indicate that incorporating external knowledge into the embedding can enhance the performance of the FTR task.

\begin{table}[]
\centering
\caption{Results of different function type recovery methods.}
\label{tab:functype}
\begin{tabular}{c|cc}
\toprule[1pt]
Model              & Train Accuracy & Test Accuracy \\ \hline
EKLAVYA (word2vec) & 0.987             & 0.809            \\
PalmTree           & 0.946             & 0.808            \\
\sysname-zero        & 0.939             & 0.856            \\
\sysname             & 0.948             & 0.876            \\ \bottomrule[1pt]
\end{tabular}
\end{table}

\begin{table}[t]
\centering
\caption{Results of different models for indirect call recognition (Poolsize=2).}
\label{tab:icall2}
\begin{tabular}{c|cc}
\toprule[1pt]
Model       & MRR   & Recall@1 \\ \hline
CALLEE (doc2vec)      & 0.777 & 0.553    \\
word2vec    & 0.754 & 0.540     \\
PalmTree    & 0.765 & 0.530     \\
\sysname-zero & 0.790  & 0.580     \\
\sysname      & 0.808 & 0.616    \\ \bottomrule[1pt]
\end{tabular}
\end{table}

\subsubsection{Indirect Call Recognition}


Indirect Call Recognition (ICR) aims to identify potential callees of indirect function calls in binary code.

We compare \sysname, PalmTree, and CALLEE in this task. CALLEE uses doc2vec~\cite{le2014doc2vec} as the assembly sequence embedding method and performs callsite-callee matching with a Siamese neural network. Therefore, we replace the assembly sequence embedding method with \sysname, word2vec, and PalmTree for comparison.

We construct a dataset for the ICR task based on SPEC CPU 2006. By running tests of SPEC CPU 2006 and using dynamic instrumentation techniques, we record the program's runtime behaviors, including the ground truth for indirect calls. In total, we collected 52,062 indirect calls from 21 projects.

Instead of using Precision/Recall/F1 of binary classification, we adopt the following two more comprehensive metrics:
\[
\text{Recall}@k = \frac{1}{|\mathcal{F}_{\text{callee}}|} \sum_{f_{i}\in \mathcal{F}_{\text{callee}}}\mathbb{I}(\text{Rank}_{f^{gt}_{i}} \le k)
\]
\[
\text{MRR} = \frac{1}{|\mathcal{F}_{\text{callee}}|} \sum_{f_{i} \in \mathcal{F}_{\text{callee}}} \frac{1}{Rank_{f^{gt}_{i}}}
\]
where $\mathcal{F}_{\text{callee}}$ is a callee pool, $\mathcal{G}_{\text{callee}}$ is the ground truth callee pool,
$\mathbb{I}$ is the indicator function.
And we denote a query callee as $f_{i} \in \mathcal{F}_{\text{callee}}$ and its corresponding ground truth function as $ f^{gt}_{i} \in \mathcal{G}_{\text{callee}}$.

\begin{table}[t]
\centering
\caption{Results of different models for indirect call recognition (Poolsize=32).}
\label{tab:icall32}
\begin{tabular}{c|cc}
\toprule[1pt]
Model       & MRR   & Recall@1 \\ \hline
CALLEE (doc2vec)     & 0.172 & 0.057    \\
word2vec    & 0.155 & 0.051    \\
PalmTree    & 0.154 & 0.046    \\
\sysname-zero & 0.231 & 0.094    \\
\sysname      & 0.280  & 0.133    \\ \bottomrule[1pt]
\end{tabular}
\end{table}

\begin{figure}[]
  \centering
  \includegraphics[width=\linewidth]{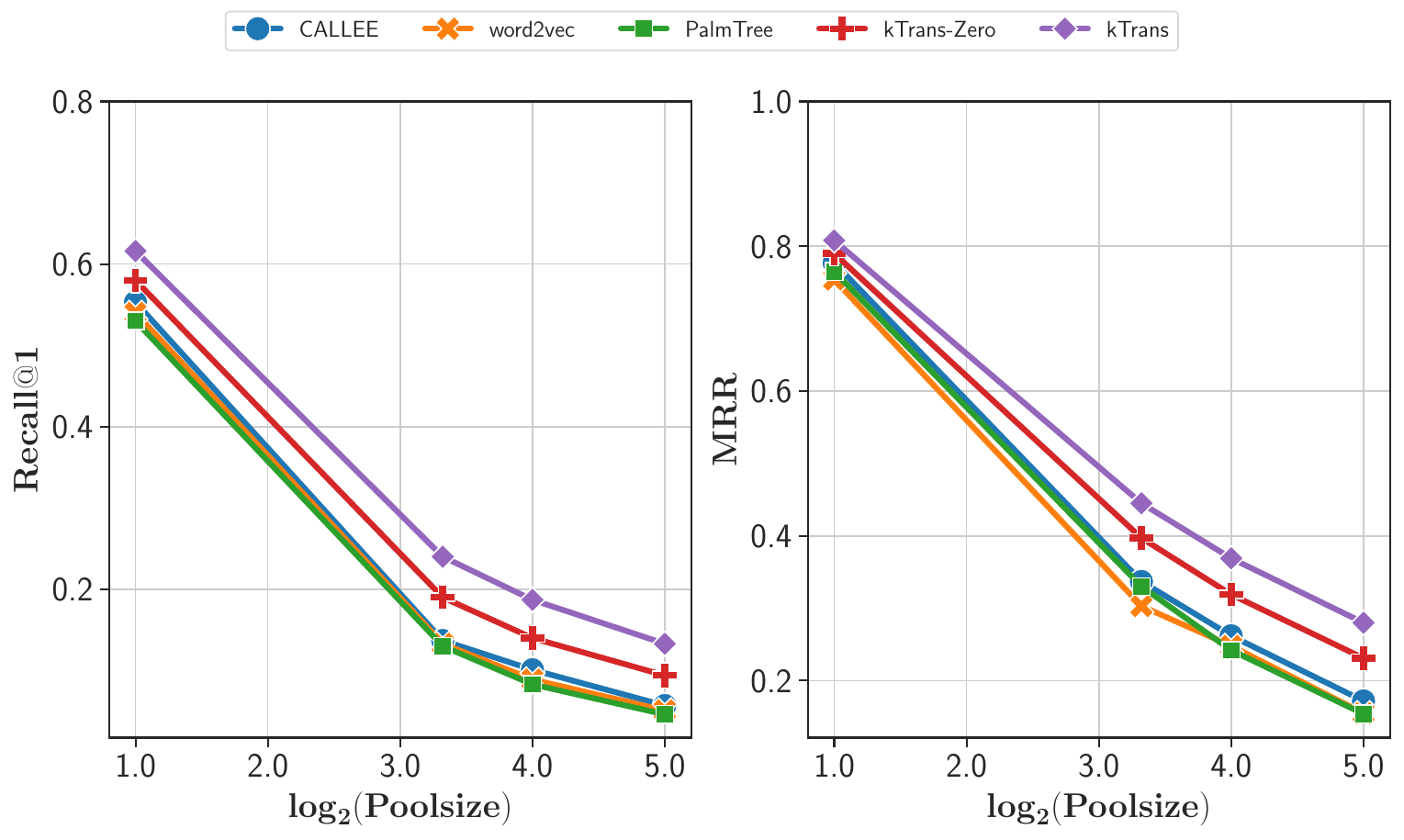}
  \caption{The performance of different models for indirect call recognition with respect to pool size.}
  \label{fig:icall}
\end{figure}

As shown in Table~\ref{tab:icall2} and Table~\ref{tab:icall32}, \sysname achieves the best performance with pool sizes of 2 and 32. We also examine the relationship between MRR, Recall@1, and pool size for each model. As depicted in Figure~\ref{fig:icall}, as the pool size increases, there is a decline in MRR and Recall@1 for all models, but \sysname maintains its superiority over other models.

Therefore, these results indicate that incorporating external knowledge into the embedding can enhance the performance of the ICR task.

In conclusion, \sysname can effectively improve the performance of downstream tasks such as BCSD, FTR, and ICR.

\section{Future Directions for Binary Code Embedding}

\subsection{Larger Domain Models}
As shown in \ref{eval:RQ4}, the current Transformer model still has room for improvement in difficult analysis tasks such as ICR.
According to the scaling law~\cite{deepmind2022scalinglaw} of Transformers, large language models scale better with bigger models and more data. 
Thus by leveraging the advancements in pre-training techniques, such as AMP~\cite{micikevicius2017mixedprecision} and DeepSpeed~\cite{rasley2020deepspeed}, we can create more powerful models specifically tailored for binary code. 
These larger domain models can potentially capture more intricate patterns and dependencies within the binary code, leading to improved performance in various tasks. 


\subsection{Cost-Effective Models}
Training and running large-scale language models can be computationally expensive and resource-intensive. Therefore, exploring cost-effective techniques for binary code embedding is an important research direction.
One possible approach is to investigate methods for transfer learning, where a pre-trained model on a large general-purpose dataset is fine-tuned on a smaller binary-specific dataset. By leveraging the knowledge learned from the general domain, the model can potentially achieve better performance with fewer computational resources. 
Another approach is to explore teacher-student paradigm for efficient model training.
This method involves training a smaller, simpler model (the student) to learn from a larger, more complex model (the teacher), to obtain better performance for the student model with fewer computational resources.
These cost-effective training methods will make the development and deployment of binary code embedding models more accessible to researchers and practitioners.

\subsection{Combination with General Large Language Models}
General large language models, such as GPT-4~\cite{openai2023gpt4}, have demonstrated remarkable capabilities in natural language understanding and generation. Combining these general large language models with binary code embedding can lead to even more powerful models for understanding and analyzing binary code. By leveraging the strengths of both types of models, we can benefit from the contextual understanding and language modeling capabilities of large language models, while also incorporating the specific domain knowledge encoded in the binary code embedding. This combination can enable a broader range of applications, such as code synthesis, vulnerability detection, and program comprehension. 
However, it is important to address the challenges of aligning the specific domain knowledge of binary code with the more general knowledge of large language models, ensuring accurate and reliable results in the binary code context.

\section{Conclusion}

In this paper, we propose \sysname, a binary code embedding model incorporating external knowledge. We leverage Transformer-based architectures to capture the semantics of assembly instructions and utilize knowledge injection techniques to enhance the model's performance. Through extensive evaluation on diverse tasks and datasets, \sysname consistently outperformed baseline models, showcasing its effectiveness in assembly language understanding. The results demonstrated superior performance in  binary code similarity detection, function type recovery, and indirect call recognition tasks. 
The findings suggest that injecting external knowledge into assembly language models significantly improves their capabilities. Furthermore, we discuss future research directions including exploring larger domain models, cost-effective models, and combination with general large language models. By addressing these directions, we can further promote assembly language modeling for software security and binary code analysis.



\section*{Availability}

\sysname is publicly \href{https://github.com/Learner0x5a/kTrans-release}{available}.


\bibliographystyle{plain}
\bibliography{ref}

\begin{thebibliography}{10}

\bibitem{Ahmad2021UnifiedPF}
Wasi~Uddin Ahmad, Saikat Chakraborty, Baishakhi Ray, and Kai-Wei Chang.
\newblock Unified pre-training for program understanding and generation.
\newblock {\em ArXiv}, abs/2103.06333, 2021.

\bibitem{Allamanis2017LearningTR}
Miltiadis Allamanis, Marc Brockschmidt, and Mahmoud Khademi.
\newblock Learning to represent programs with graphs.
\newblock {\em ArXiv}, abs/1711.00740, 2017.

\bibitem{alon2018code2seq}
Uri Alon, Shaked Brody, Omer Levy, and Eran Yahav.
\newblock code2seq: Generating sequences from structured representations of
  code.
\newblock {\em arXiv preprint arXiv:1808.01400}, 2018.

\bibitem{alon2019code2vec}
Uri Alon, Meital Zilberstein, Omer Levy, and Eran Yahav.
\newblock code2vec: Learning distributed representations of code.
\newblock {\em Proceedings of the ACM on Programming Languages}, 3(POPL):1--29,
  2019.

\bibitem{artuso2022binbert}
Fiorella Artuso, Marco Mormando, Giuseppe~A Di~Luna, and Leonardo Querzoni.
\newblock Binbert: Binary code understanding with a fine-tunable and
  execution-aware transformer.
\newblock {\em arXiv preprint arXiv:2208.06692}, 2022.

\bibitem{austin2021program}
Jacob Austin, Augustus Odena, Maxwell Nye, Maarten Bosma, Henryk Michalewski,
  David Dohan, Ellen Jiang, Carrie Cai, Michael Terry, Quoc Le, et~al.
\newblock Program synthesis with large language models.
\newblock {\em arXiv preprint arXiv:2108.07732}, 2021.

\bibitem{NEURIPS2020_1457c0d6}
Tom Brown, Benjamin Mann, Nick Ryder, Melanie Subbiah, Jared~D Kaplan, Prafulla
  Dhariwal, Arvind Neelakantan, Pranav Shyam, Girish Sastry, Amanda Askell,
  Sandhini Agarwal, Ariel Herbert-Voss, Gretchen Krueger, Tom Henighan, Rewon
  Child, Aditya Ramesh, Daniel Ziegler, Jeffrey Wu, Clemens Winter, Chris
  Hesse, Mark Chen, Eric Sigler, Mateusz Litwin, Scott Gray, Benjamin Chess,
  Jack Clark, Christopher Berner, Sam McCandlish, Alec Radford, Ilya Sutskever,
  and Dario Amodei.
\newblock Language models are few-shot learners.
\newblock In H.~Larochelle, M.~Ranzato, R.~Hadsell, M.F. Balcan, and H.~Lin,
  editors, {\em Advances in Neural Information Processing Systems}, volume~33,
  pages 1877--1901. Curran Associates, Inc., 2020.

\bibitem{chua2017type}
Zheng~Leong Chua, Shiqi Shen, Prateek Saxena, and Zhenkai Liang.
\newblock Neural nets can learn function type signatures from binaries.
\newblock In {\em USENIX Security Symposium}, pages 99--116, 2017.

\bibitem{chung2014GRU}
Junyoung Chung, Caglar Gulcehre, KyungHyun Cho, and Yoshua Bengio.
\newblock Empirical evaluation of gated recurrent neural networks on sequence
  modeling, 2014.

\bibitem{devlin2018bert}
Jacob Devlin, Ming-Wei Chang, Kenton Lee, and Kristina Toutanova.
\newblock Bert: Pre-training of deep bidirectional transformers for language
  understanding.
\newblock {\em arXiv preprint arXiv:1810.04805}, 2018.

\bibitem{ding2019asm2vec}
Steven~HH Ding, Benjamin~CM Fung, and Philippe Charland.
\newblock Asm2vec: Boosting static representation robustness for binary clone
  search against code obfuscation and compiler optimization.
\newblock In {\em 2019 IEEE Symposium on Security and Privacy (SP)}, pages
  472--489. IEEE, 2019.

\bibitem{dullien2005graph}
Thomas Dullien and Rolf Rolles.
\newblock Graph-based comparison of executable objects (english version).
\newblock {\em Sstic}, 5(1):3, 2005.

\bibitem{feng2016scalable}
Qian Feng, Rundong Zhou, Chengcheng Xu, Yao Cheng, Brian Testa, and Heng Yin.
\newblock Scalable graph-based bug search for firmware images.
\newblock In {\em Proceedings of the 2016 ACM SIGSAC Conference on Computer and
  Communications Security}, pages 480--491, 2016.

\bibitem{feng-etal-2020-codebert}
Zhangyin Feng, Daya Guo, Duyu Tang, Nan Duan, Xiaocheng Feng, Ming Gong, Linjun
  Shou, Bing Qin, Ting Liu, Daxin Jiang, and Ming Zhou.
\newblock {C}ode{BERT}: A pre-trained model for programming and natural
  languages.
\newblock In {\em Findings of the Association for Computational Linguistics:
  EMNLP 2020}, pages 1536--1547, Online, November 2020. Association for
  Computational Linguistics.

\bibitem{gao2008binhunt}
Debin Gao, Michael~K Reiter, and Dawn Song.
\newblock Binhunt: Automatically finding semantic differences in binary
  programs.
\newblock In {\em International Conference on Information and Communications
  Security}, pages 238--255. Springer, 2008.

\bibitem{gao2018vulseeker}
Jian Gao, Xin Yang, Ying Fu, Yu~Jiang, and Jiaguang Sun.
\newblock Vulseeker: a semantic learning based vulnerability seeker for
  cross-platform binary.
\newblock In {\em 2018 33rd IEEE/ACM International Conference on Automated
  Software Engineering (ASE)}, pages 896--899. IEEE, 2018.

\bibitem{guo2019deepvsa}
Wenbo Guo, Dongliang Mu, Xinyu Xing, Min Du, and Dawn Song.
\newblock Deepvsa: Facilitating value-set analysis with deep learning for
  postmortem program analysis.
\newblock In {\em USENIX Security Symposium}, pages 1787--1804, 2019.

\bibitem{he2022binprov}
Xu~He, Shu Wang, Yunlong Xing, Pengbin Feng, Haining Wang, Qi~Li, Songqing
  Chen, and Kun Sun.
\newblock Binprov: Binary code provenance identification without disassembly.
\newblock In {\em Proceedings of the 25th International Symposium on Research
  in Attacks, Intrusions and Defenses}, pages 350--363, 2022.

\bibitem{IDAPro}
{Hex-Rays SA}.
\newblock {IDA Pro: a cross-platform multi-processor disassembler and
  debugger.}
\newblock \url{http://www.hex-rays.com/products/ida/index.shtml}.

\bibitem{LSTM}
Sepp Hochreiter and Jürgen Schmidhuber.
\newblock Long short-term memory.
\newblock {\em Neural Computation}, 9(8):1735--1780, 1997.

\bibitem{deepmind2022scalinglaw}
Jordan Hoffmann, Sebastian Borgeaud, Arthur Mensch, Elena Buchatskaya, Trevor
  Cai, Eliza Rutherford, Diego de~Las Casas, Lisa~Anne Hendricks, Johannes
  Welbl, Aidan Clark, et~al.
\newblock Training compute-optimal large language models.
\newblock {\em arXiv preprint arXiv:2203.15556}, 2022.

\bibitem{hu2009SMIT}
Xin Hu, Tzi-cker Chiueh, and Kang~G Shin.
\newblock Large-scale malware indexing using function-call graphs.
\newblock In {\em Proceedings of the 16th ACM conference on Computer and
  communications security}, pages 611--620, 2009.

\bibitem{huang2021makes}
Yu~Huang, Chenzhuang Du, Zihui Xue, Xuanyao Chen, Hang Zhao, and Longbo Huang.
\newblock What makes multi-modal learning better than single (provably).
\newblock {\em Advances in Neural Information Processing Systems},
  34:10944--10956, 2021.

\bibitem{iyer-etal-2016-summarizing}
Srinivasan Iyer, Ioannis Konstas, Alvin Cheung, and Luke Zettlemoyer.
\newblock Summarizing source code using a neural attention model.
\newblock In {\em Proceedings of the 54th Annual Meeting of the Association for
  Computational Linguistics (Volume 1: Long Papers)}, pages 2073--2083, Berlin,
  Germany, August 2016. Association for Computational Linguistics.

\bibitem{kim:tse:2022}
Dongkwan Kim, Eunsoo Kim, Sang~Kil Cha, Sooel Son, and Yongdae Kim.
\newblock Revisiting binary code similarity analysis using interpretable
  feature engineering and lessons learned.
\newblock {\em IEEE Transactions on Software Engineering}, pages 1--23, 2022.

\bibitem{kong2019mutual}
Lingpeng Kong, Cyprien de~Masson d'Autume, Wang Ling, Lei Yu, Zihang Dai, and
  Dani Yogatama.
\newblock A mutual information maximization perspective of language
  representation learning.
\newblock {\em arXiv preprint arXiv:1910.08350}, 2019.

\bibitem{le2014doc2vec}
Quoc Le and Tomas Mikolov.
\newblock Distributed representations of sentences and documents.
\newblock In {\em International conference on machine learning}, pages
  1188--1196. PMLR, 2014.

\bibitem{lee2017instruction2vec}
Young~Jun Lee, Sang-Hoon Choi, Chulwoo Kim, Seung-Ho Lim, and Ki-Woong Park.
\newblock Learning binary code with deep learning to detect software weakness.
\newblock In {\em KSII the 9th international conference on internet (ICONI)
  2017 symposium}, 2017.

\bibitem{li2021palmtree}
Xuezixiang Li, Yu~Qu, and Heng Yin.
\newblock Palmtree: Learning an assembly language model for instruction
  embedding.
\newblock In {\em Proceedings of the 2021 ACM SIGSAC Conference on Computer and
  Communications Security}, pages 3236--3251, 2021.

\bibitem{liu2018alphadiff}
Bingchang Liu, Wei Huo, Chao Zhang, Wenchao Li, Feng Li, Aihua Piao, and Wei
  Zou.
\newblock $\alpha$diff: cross-version binary code similarity detection with
  dnn.
\newblock In {\em Proceedings of the 33rd ACM/IEEE International Conference on
  Automated Software Engineering}, pages 667--678, 2018.

\bibitem{liu2019roberta}
Yinhan Liu, Myle Ott, Naman Goyal, Jingfei Du, Mandar Joshi, Danqi Chen, Omer
  Levy, Mike Lewis, Luke Zettlemoyer, and Veselin Stoyanov.
\newblock Roberta: A robustly optimized bert pretraining approach.
\newblock {\em arXiv preprint arXiv:1907.11692}, 2019.

\bibitem{vulhawk}
Zhenhao Luo, Pengfei Wang, Baosheng Wang, Yong Tang, Wei Xie, Xu~Zhou, Danjun
  Liu, and Kai Lu.
\newblock Vulhawk: Cross-architecture vulnerability detection with
  entropy-based binary code search.
\newblock In {\em 30th Annual Network and Distributed System Security
  Symposium, {NDSS} 2023, San Diego, California, USA, February 27 - March 3,
  2023}. The Internet Society, 2023.

\bibitem{marcelli2022cisco}
Andrea Marcelli, Mariano Graziano, Xabier Ugarte-Pedrero, Yanick Fratantonio,
  Mohamad Mansouri, and Davide Balzarotti.
\newblock How machine learning is solving the binary function similarity
  problem.
\newblock In {\em 31st USENIX Security Symposium (USENIX Security 22)}, pages
  2099--2116, 2022.

\bibitem{massarelli2019investigating}
Luca Massarelli, Giuseppe~A Di~Luna, Fabio Petroni, Leonardo Querzoni, and
  Roberto Baldoni.
\newblock Investigating graph embedding neural networks with unsupervised
  features extraction for binary analysis.
\newblock In {\em Proceedings of the 2nd Workshop on Binary Analysis Research
  (BAR)}, 2019.

\bibitem{massarelli2019safe}
Luca Massarelli, Giuseppe~Antonio Di~Luna, Fabio Petroni, Roberto Baldoni, and
  Leonardo Querzoni.
\newblock Safe: Self-attentive function embeddings for binary similarity.
\newblock In {\em International Conference on Detection of Intrusions and
  Malware, and Vulnerability Assessment}, pages 309--329. Springer, 2019.

\bibitem{micikevicius2017mixedprecision}
Paulius Micikevicius, Sharan Narang, Jonah Alben, Gregory Diamos, Erich Elsen,
  David Garcia, Boris Ginsburg, Michael Houston, Oleksii Kuchaiev, Ganesh
  Venkatesh, et~al.
\newblock Mixed precision training.
\newblock {\em arXiv preprint arXiv:1710.03740}, 2017.

\bibitem{mikolov2013word2vec}
Tomas Mikolov, Kai Chen, Greg Corrado, and Jeffrey Dean.
\newblock Efficient estimation of word representations in vector space.
\newblock {\em arXiv preprint arXiv:1301.3781}, 2013.

\bibitem{Capstone}
{Nguyen Anh Quynh}.
\newblock {Capstone: Next-gen disassembly framework.}

\bibitem{openai2023gpt4}
OpenAI.
\newblock Gpt-4 technical report, 2023.

\bibitem{otsubo2020glassesx}
Yuhei Otsubo, Akira Otsuka, Mamoru Mimura, Takeshi Sakaki, and Hiroshi Ukegawa.
\newblock o-glassesx: Compiler provenance recovery with attention mechanism
  from a short code fragment.
\newblock In {\em Proceedings of the 3nd Workshop on Binary Analysis Research},
  2020.

\bibitem{pei2020xda}
Kexin Pei, Jonas Guan, David Williams-King, Junfeng Yang, and Suman Jana.
\newblock Xda: Accurate, robust disassembly with transfer learning.
\newblock {\em arXiv preprint arXiv:2010.00770}, 2020.

\bibitem{pennington2014glove}
Jeffrey Pennington, Richard Socher, and Christopher~D Manning.
\newblock Glove: Global vectors for word representation.
\newblock In {\em Proceedings of the 2014 conference on empirical methods in
  natural language processing (EMNLP)}, pages 1532--1543, 2014.

\bibitem{peters-etal-2018-deep}
Matthew~E. Peters, Mark Neumann, Mohit Iyyer, Matt Gardner, Christopher Clark,
  Kenton Lee, and Luke Zettlemoyer.
\newblock Deep contextualized word representations.
\newblock In {\em Proceedings of the 2018 Conference of the North {A}merican
  Chapter of the Association for Computational Linguistics: Human Language
  Technologies, Volume 1 (Long Papers)}, pages 2227--2237, New Orleans,
  Louisiana, June 2018. Association for Computational Linguistics.

\bibitem{Pradel2018DeepBugsAL}
Michael Pradel and Koushik Sen.
\newblock Deepbugs: a learning approach to name-based bug detection.
\newblock {\em Proceedings of the ACM on Programming Languages}, 2:1 -- 25,
  2018.

\bibitem{radfordimproving}
Alec Radford, Karthik Narasimhan, Tim Salimans, and Ilya Sutskever.
\newblock Improving language understanding by generative pre-training.

\bibitem{radford2019language}
Alec Radford, Jeffrey Wu, Rewon Child, David Luan, Dario Amodei, Ilya
  Sutskever, et~al.
\newblock Language models are unsupervised multitask learners.
\newblock {\em OpenAI blog}, 1(8):9, 2019.

\bibitem{rasley2020deepspeed}
Jeff Rasley, Samyam Rajbhandari, Olatunji Ruwase, and Yuxiong He.
\newblock Deepspeed: System optimizations enable training deep learning models
  with over 100 billion parameters.
\newblock In {\em Proceedings of the 26th ACM SIGKDD International Conference
  on Knowledge Discovery \& Data Mining}, pages 3505--3506, 2020.

\bibitem{shin2015boundary}
Eui Chul~Richard Shin, Dawn Song, and Reza Moazzezi.
\newblock Recognizing functions in binaries with neural networks.
\newblock In {\em 24th $\{$USENIX$\}$ Security Symposium ($\{$USENIX$\}$
  Security 15)}, pages 611--626, 2015.

\bibitem{svyatkovskiy2019pythia}
Alexey Svyatkovskiy, Ying Zhao, Shengyu Fu, and Neel Sundaresan.
\newblock Pythia: Ai-assisted code completion system.
\newblock In {\em Proceedings of the 25th ACM SIGKDD international conference
  on knowledge discovery \& data mining}, pages 2727--2735, 2019.

\bibitem{wang2022jtrans}
Hao Wang, Wenjie Qu, Gilad Katz, Wenyu Zhu, Zeyu Gao, Han Qiu, Jianwei Zhuge,
  and Chao Zhang.
\newblock jtrans: jump-aware transformer for binary code similarity detection.
\newblock In {\em Proceedings of the 31st ACM SIGSOFT International Symposium
  on Software Testing and Analysis}, pages 1--13, 2022.

\bibitem{xu2018deeprefiner}
Ke~Xu, Yingjiu Li, Robert~H Deng, and Kai Chen.
\newblock Deeprefiner: Multi-layer android malware detection system applying
  deep neural networks.
\newblock In {\em 2018 IEEE European Symposium on Security and Privacy
  (EuroS\&P)}, pages 473--487. IEEE, 2018.

\bibitem{xu2017neural}
Xiaojun Xu, Chang Liu, Qian Feng, Heng Yin, Le~Song, and Dawn Song.
\newblock Neural network-based graph embedding for cross-platform binary code
  similarity detection.
\newblock In {\em Proceedings of the 2017 ACM SIGSAC Conference on Computer and
  Communications Security}, pages 363--376, 2017.

\bibitem{yu2020order}
Zeping Yu, Rui Cao, Qiyi Tang, Sen Nie, Junzhou Huang, and Shi Wu.
\newblock Order matters: Semantic-aware neural networks for binary code
  similarity detection.
\newblock In {\em Proceedings of the AAAI Conference on Artificial
  Intelligence}, volume~34, pages 1145--1152, 2020.

\bibitem{zhang2022combo}
Yifan Zhang, Chen Huang, Yueke Zhang, Kevin Cao, Scott~Thomas Andersen, Huajie
  Shao, Kevin Leach, and Yu~Huang.
\newblock Combo: Pre-training representations of binary code using contrastive
  learning.
\newblock {\em arXiv preprint arXiv:2210.05102}, 2022.

\bibitem{zhu2023callee}
Wenyu Zhu, Zhiyao Feng, Zihan Zhang, Jianjun Chen, Zhijian Ou, Min Yang, and
  Chao Zhang.
\newblock Callee: Recovering call graphs for binaries with transfer and
  contrastive learning.
\newblock In {\em 2023 IEEE Symposium on Security and Privacy (SP)}, pages
  1953--1970. IEEE Computer Society, 2023.

\end{thebibliography}

\end{document}